\documentclass[10pt, journal, comsoc]{IEEEtran}

\ifCLASSINFOpdf
  \usepackage[pdftex]{graphicx}
  \DeclareGraphicsExtensions{.pdf,.jpeg,.png}
\else
  \usepackage[dvips]{graphicx}
  \DeclareGraphicsExtensions{.eps}
\fi

\usepackage[english]{babel}										%
\usepackage[protrusion=true,expansion=true]{microtype}				%
\usepackage{amsmath,amsfonts}					%
\usepackage{graphicx}									%
\usepackage[svgnames]{xcolor}									%
\usepackage{epstopdf}												%
\usepackage{booktabs}												%
\usepackage[square,sort,comma,numbers,compress]{natbib}
\usepackage[ruled,vlined,commentsnumbered,linesnumbered,lined,boxed]{algorithm2e}
\usepackage{algorithmicx}
\usepackage{algpseudocode}

\usepackage[hang, small, labelfont=bf, textfont=it, up]{caption}	%
\usepackage[skip=0cm,list=true]{subcaption}
\usepackage{paralist}
\usepackage{balance}

\algrenewcommand\algorithmicindent{1.0em}%

\let\oldenumerate\enumerate
\renewcommand{\enumerate}{
  \oldenumerate
  \setlength{\itemsep}{1pt}
  \setlength{\parskip}{0pt}
  \setlength{\parsep}{1pt}
}

\let\olditemize\itemize
\renewcommand{\itemize}{
  \olditemize
  \setlength{\itemsep}{1pt}
  \setlength{\parskip}{0pt}
  \setlength{\parsep}{1pt}
}

\title{The Switch from Conventional to SDN: The Case for Transport-Agnostic Congestion Control}					

\author{
   Ahmed M. Abdelmoniem and %
   Brahim Bensaou%
   \thanks{Ahmed M. Abdelmoniem is currently with School of EECS, Queen Mary University of London and CS Department, Assiut University, Egypt.
E-mail: ahmed.sayed@qmul.ac.uk
}
\thanks{Brahim Bensaou is currently with Department of Computer Science and Engineering, HKUST, Hong Kong SAR, PRC. 
 }%
 \thanks{This work combines works published in IEEE ICC 2016~\cite{Ahmed-ICC-2016-1} and IEEE LCN 2017~\cite{Ahmed-LCN-2017}.
 }
}

\begin{document}

\newcommand{\SWITCH}[1]{\State \textbf{switch} (#1)}
\newcommand{\ENDSWITCH}{\State \textbf{end switch}}
\newcommand{\CASE}[1]{\State \textbf{case} #1\textbf{:} \begin{ALC@g}}
\newcommand{\ENDCASE}{\end{ALC@g}}
\newcommand{\CASELINE}[1]{\State \textbf{case} #1\textbf{:} }
\newcommand{\DEFAULT}{\State \textbf{default:} \begin{ALC@g}}
\newcommand{\ENDDEFAULT}{\end{ALC@g}}
\newcommand{\DEFAULTLINE}[1]{\State \textbf{default:} }

\maketitle

\begin{abstract}
To meet the timing requirements of interactive applications, the no-frills congestion-agnostic transport protocols like UDP are increasingly deployed side-by-side in the same network with congestion-responsive TCP. In cloud platforms, even though the computation and storage is totally virtualized, they lack a true virtualization mechanism for the network (i.e., the underlying data centers networks). The impact of such lack of isolation services, may result into frequent outages (for some applications) when such diverse traffics contend for the small buffers in the commodity switches used in data centers. In this paper, we explore the design space of a simple, practical and \textbf{transport-agnostic} scheme to enable a scalable and flexible end-to-end congestion control in data centers. Then, we present the the shortcomings of coupling the monitoring and control of congestion in the conventional system and discuss how a Software-Defined Network (SDN) would provide an appealing alternative to circumvent the problems of the conventional system. The two systems implements a software-based congestion control mechanisms that perform monitoring, control decisions and traffic control enforcement functions. Both systems are designed with a major assumption that the applications (or transport protocols) are non-cooperative with the system, ultimately making it deployable in existing data centers without any service disruption or hardware upgrade. Both systems are implemented and evaluated via simulation in NS2 as well as real-life small-scale test-bed deployment and experiments. %

\end{abstract}


\section{Introduction}
\label{sec:problem}
To achieve isolation among tenants and use resources more efficiently, resource virtualization has become a common practice in today's public data centers. In most cases, each tenant is provisioned with virtual machines assigned with dedicated virtual CPU cores, memory, storage, and a virtual network interface card (NIC) that sends traffic over the underlying shared physical NIC. Typically, tenants can not assume predictability nor measurability of bounds on network performance, as no mechanisms are deployed to explicitly allocate and enforce bandwidth in the cloud. Nevertheless, cloud operators can provide tenants with better virtual network management thanks to the recent developments in control plane functions. For example, Amazon introduced ``Virtual Private Cloud (VPC)'' \cite{AWS-VPC} to allow easy creation and management of tenant's private virtual network. VPC can be viewed as an abstraction layer running on top of the non-isolated, shared network resources of AWS' public cloud. Additionally, Software Defined Networking (SDN) \cite{SDN} has been effectively deployed to drive inter- and intra-data center communications with added features to make the virtualization and other network aspects easy to manage. For example, both Google \cite{Jain2013} and Microsoft \cite{Hong2013} have deployed fully operational SDN-based WAN networks to support standard routing protocols as well as centralized traffic engineering between their data centers. 

In contrast, the data plane in intra-datacenter networks has seen little progress in managing bandwidth to overcome congestion, improve efficiency, and apportioning it adequately to provide isolation between competing tenants to meet their target performance requirements. In principle, isolation can simply be achieved through static reservation \cite{Guo2010, Ballani2011a, Ahmed-CONEXT-2020, Ahmed-SIDCo-MLSys21, Ahmed-AQFL-21}, where tenants can enjoy a predictable, congestion-free network performance. However, static reservations lead to inefficient utilization of the network capacity. To avoid such pitfall, tenants should be assigned a minimum bandwidth using the so-called hose model \cite{Duffield1999} which abstracts the network between the VMs of one tenant as a single virtual switch (vSwitch). In such setup, different VMs may reside on any physical machine in the datacenter, yet, each VM should be able to send traffic at the full virtual port rate as determined by the vSwitch abstraction layer. Such VMs should enjoy the allocated rate regardless of the traffic patterns of co-existing VMs and/or the nature of the workload generated by competing VMs. %

The following are the necessary elements that can be incorporated together for this purpose: 
\begin{itemize}
\item An intelligent and scalable VM admission mechanism within the datacenter for VM placement where minimum bandwidth is available. To facilitate this, topologies with bottlenecks at the core switches (such as uplink over-subscription or a low bisection bandwidth) should be avoided if possible.
\item A methodology to fully utilize the available high bisection bandwidth (e.g., a load balancing mechanism and/or multi-path transport/routing protocols). %
\item A rate adaptation technique to ensure conformance of VM sending rates to their allocated bandwidth, while penalizing misbehaving ones. 
\end{itemize}

\begin{figure}[t]
	\centering	 
	 \begin{subfigure}[ht]{0.9\columnwidth}
                \includegraphics[width=\textwidth, height=5cm]{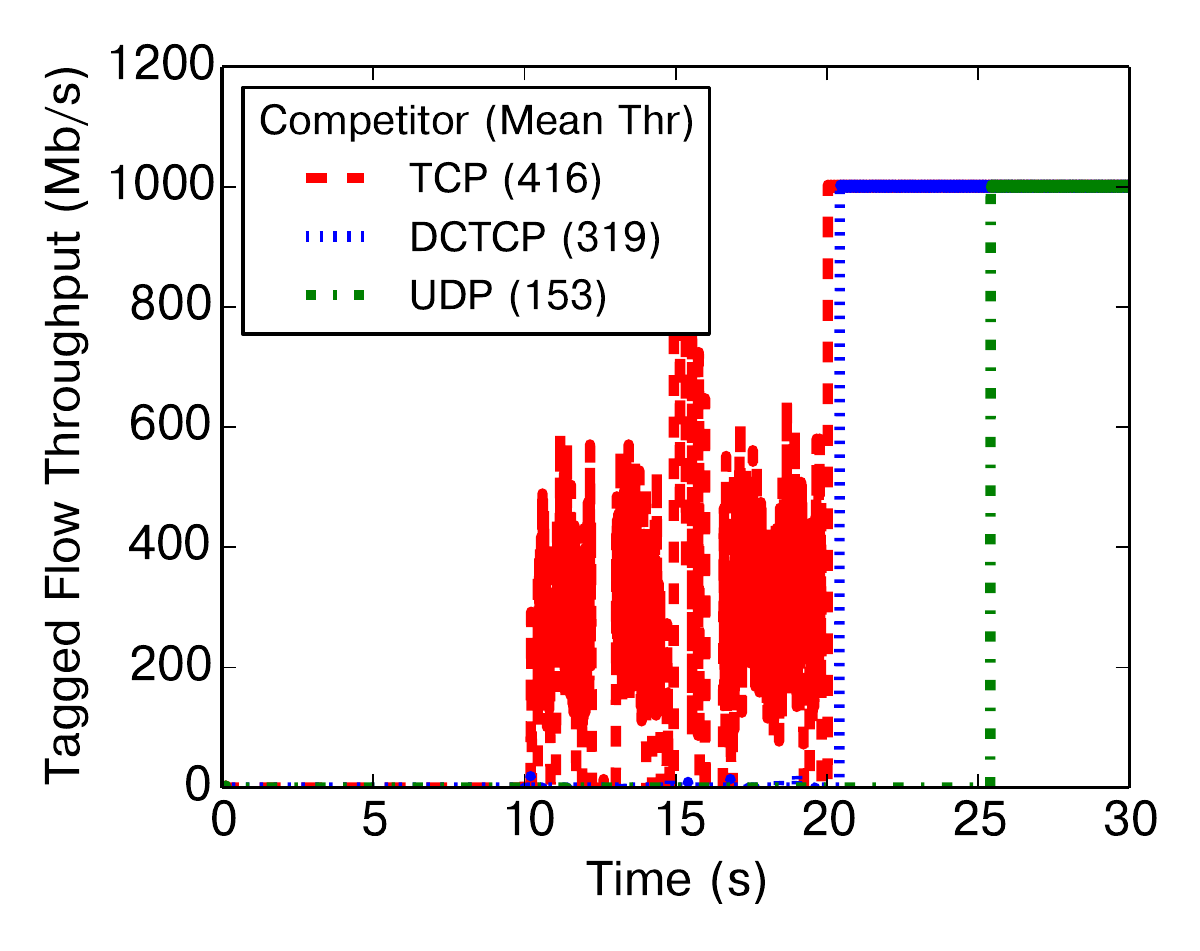}
                \caption{ECN disabled}
                \label{fig:noecn}							
        \end{subfigure}
			  \hfill
        \begin{subfigure}[ht]{0.9\columnwidth}
				\includegraphics[width=\textwidth, height=5cm]{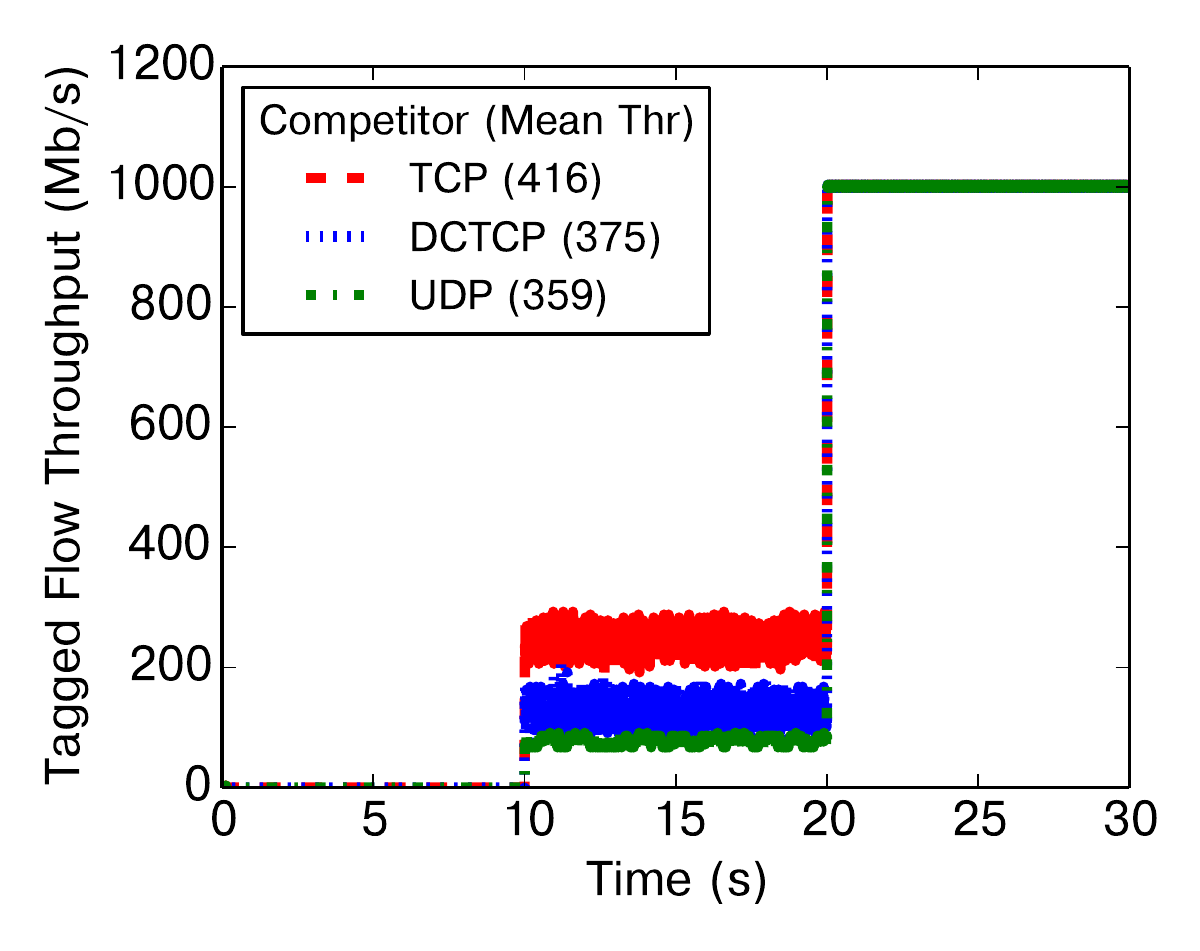}
                \caption{ECN enabled}
                \label{fig:ecn}
        \end{subfigure}
 \caption{The instantaneous and mean goodput of the tagged TCP flow (in the legends) while competing with 3 TCP, DCTCP or UDP senders. The link capacity is 1Gbps. In interval [0,10] only the competitors are active, in [10,20] all the flows are active and in [20,30] only the tagged TCP flow is active.}
	\label{fig:unfairness}
\end{figure}

A number of interesting research works have investigated more or less successfully the first two elements of this framework \cite{Al-Fares2008, Greenberg2009, Benson2011, Raiciu2011, Ahmed-ICDCS-2019-1, Ahmed-ICDCS-2019-2,  Ahmed-INFOCOM-2018, Ahmed-INFOCOM-2019,  Ahmed-ICC-2019, Ahmed-urnas-2019, Ahmed-DC2-INFOCOM21, Ahmed-ICPP-2020, Ahmed-TON-2021}. In \cite{Al-Fares2008, Greenberg2009}, highly scalable network topologies offering a 1:1 over-subscription and a high bisection bandwidth were proposed. These topologies are shown to be easily deployable in practice and can simplify the VM placement at any physical machine with sufficient bandwidth to support the VM. Efficient routing and transport protocols \cite{Benson2011, Raiciu2011} were designed for DCN to achieve a high utilization of the available capacity. Finally, in terms of traffic control, much of the recent work \cite{Alizadeh2010, Wu2013} focused on restructuring TCP congestion control and its variants to efficiently utilize and fairly share bandwidth among flows (in homogeneous deployments). However, these techniques fall short of providing true isolation among tenants (e.g., a tenant may gain more bandwidth by opening parallel connections or by using aggressive transport protocol like UDP). It is common, in multi-tenant environments, that non-homogeneous transport protocols co-exist leading to starvation of the cooperative ones \cite{Irteza2014}. To illustrate this problem simply, we conduct a set of simulations in which we compare the performance of a tagged ECN-enabled TCP (NewReno) flow that competes head-to-head i) against another TCP flow of the same type, ii) against another flow using a TCP variant designed for data centers (i.e., DCTCP which is deployed in a number of private data centers \cite{Judd2015}), and; iii) against another congestion-agnostic transport protocol (i.e., UDP which is used in MemCacheD clusters of Facebook \cite{Nishtala2013}); the results from the three experiments are superimposed and presented on the same graphs.  Similar to what was already known from the Internet, \figurename~\ref{fig:unfairness} shows that, homogeneous TCP deployments in data centers can achieve fairness, in contrast to heterogeneous deployments. In particular, we observe in \figurename\ref{fig:noecn} that when TCP is not responsive to ECN markings, it enjoys almost $\approx0\%$ its share during the periods where it competes against DCTCP or UDP. In contrast, being responsive to  ECN markings helps TCP improve its performance but still ultimately the fairness is not achieved. In \figurename~\ref{fig:ecn}, we observe that even when TCP is ECN-enabled it still loses $\approx60\%$ and $\approx72\%$ of its fair share to DCTCP and UDP, respectively.

In this paper, we rely on the SDN capability of most data center switching devices to propose a generic congestion control (SDN-GCC) mechanism to address this issue. We first introduce the idea behind SDN-GCC in Section \ref{sec:intro}, then discuss our proposed methodology and present SDN-GCC framework in Section \ref{sec:method}. We show via ns2 simulation how SDN-GCC achieves its requirements with high efficiency in Section~\ref{sec:evaluate}, then present testbed experiments in Section~\ref{sec:testbed}\footnote{Simulation and implementation code can be requested from the authors or downloaded from the following repository after they are made publicly available: http://github.com/ahmedcs/SDN-GCC}. Finally, we conclude the paper in Section \ref{sec:conclusion}.

\section{Transport Isolation Problem}
\label{sec:intro}

With the recent introduction of a significant number of new transport protocols designed for DC networks in addition to the existing protocols, the following three challenges emerged: 
\begin{inparaenum}[\itshape i) \upshape]
\item most such protocols are agnostic to the nature of the VM aggregate traffic demands leading to inefficient distribution of the network capacity among competing VMs (for instance a VM could gain more bandwidth by opening parallel TCP connections); 
\item many versions of TCP co-exist in DC networks (e.g., TCP NewReno/MacOS, compound TCP/Windows, Cubic TCP/Linux, DCTCP/Linux, and so on), leading to further inefficiency in addition to unfairness, and;
\item many DC applications rely on UDP to build custom transport protocols (e.g., \cite{Nishtala2013}), that are not responsive to congestion signals, which exacerbate the unfairness to the point of causing starvation to congestion-responsive flows.
\end{inparaenum} 
While such problems have been revealed in the context of Internet communications, two decades ago, recent studies \cite{Irteza2014, Judd2015} have confirmed that such problems of unfairness and bandwidth inefficiency also exist in DCNs despite their characteristically small delays, small buffers and different topologies from those found in the Internet. As a consequence, a new solution to the problems of congestion in DC networks is needed. Such solution must be attractive to cloud operators and cloud tenants alike. 

In particular, with the emergence of software defined networking, we see an opportunity to invoke the powerful control features and the global scope provided by SDN to revisit the problem from a different perspective, with additional realistic design constraints. As such we propose a solution with the following intuitive design requirements: 
\begin{inparaenum}[R1)]
\item \label{R1} Simplicity: to be readily deployable in existing production data centers;
\item \label{R2} Transport-agnosticism: to be effective regardless of the transport protocol;
\item \label{R3} Transparency: requires no changes to the tenant's OS (in the VM) and makes no assumption of any advanced network hardware capability other than those available in commodity SDN switches; 
\item \label{R4} Load-effectiveness: creates a minimal processing overhead on the end-host.
\end{inparaenum}

All of today's communication infrastructure from hardware devices to communication protocols have been designed with requirements derived from the global Internet. As a result to cope with scalability and AS autonomy, the decentralized approach has been adopted, relinquishing all intelligence to end systems. Yet, to enable responsiveness to congestion regardless of the transport protocol capabilities, in time-scales that commensurate with data center delays, it is preferable to adopt centralized control as it provides a global view of congestion and is known to achieve far better performance \cite{SDTCP2015, Ahmed-ICC-2017}. Nevertheless to reconcile existing hardware and protocols (designed for distributed networks) with the centralized approach, we impose design requirements R\ref{R1}-R\ref{R4} on SDN-GCC. As such the core design of SDN-GCC relies on outsourcing the congestion control decisions to the SDN controller while the enforcement of such decisions is carried out by the end-hosts hypervisors.

\section{Introduction to HyGenICC}
\label{sec:intro}

To enable responsiveness to congestion regardless of the transport protocol, one needs to return to the fundamentals and put the burden of congestion control in principle where it belongs: in the network layer. As such, in principle, such congestion control mechanism must be transparent to the transport layer protocol. However, to reconcile the principle with the practice, design requirements R\ref{R1}-R\ref{R4} must be fulfilled and thus HyGenICC outsources its congestion control building blocks to the hypervisor. 

To meet requirement R\ref{R1}, HyGenICC can be implemented either as a hypervisor-level shim-layer or as an added feature to any of the current commercial virtual switches' data-path module. The job of the added shim-layer to the hypervisor is to enforce per-VM rate control without VM cooperation nor any knowledge about its traffic patterns, workloads, or used transport protocol (TCP/UDP). %
To this end, HyGenICC maintains a \textsl{rate allocation mechanism} at each server to partition the available uplink bandwidth among VMs locally at the sending and receiving servers. In each such server, HyGenICC only needs to maintain state information per VM which meets design requirement R\ref{R4}. HyGenICC deploys a simple hypervisor-to-hypervisor (IP-to-IP) \textsl{congestion control mechanism} that relies on ECN markings (readily available in commodity switches) to infer core network congestion. HyGenICC operates at the IP level and does not interact directly with the VMs, which meets requirements R\ref{R1}, R\ref{R2} and R\ref{R3}. In addition, when detecting a highly congested path in the core network towards a destination (via ECN), HyGenICC performs admission control by refraining from accepting any further connections to this destination VM until the congestion subsides. Our design is highly scalable, responsive, work conserving and since it is IP based, it enforces the allocated bandwidth even in the presence of highly dynamic and changing traffic patterns and transport protocols. The rate allocator resolves the contention among tens-to-hundreds of co-located VMs at the servers, while the congestion control mechanism addresses the contention in the network core and pushes it back to the sources. HyGenICC also allows administrators to assign per-VM weights which directly affect the bandwidth reservation for the VMs making it appealing from cloud providers' perspective as it enables easier and more tangible bandwidth pricing and accounting.

\section{Proposed Methodology}
\label{sec:method}
First we discuss HyGenICC by imagining the datacenter network as contained within one end-host where the VMs are connected via a single virtual switch. Then, we extend this design to operate in a network of end-hosts where the datacenter fabric is treated as black box that generates congestion signals whenever congestion is experienced. 
In a single virtual switch connecting all VMs, bandwidth contention happens at the output link to the destination when multiple senders compete to send through the same output port of the virtual switch. %
The virtual switch need to distribute the available physical port's capacity among VMs and ensure compliance of the VMs with the allocated shares. Hence it needs a mechanism that detects and accounts for active VMs and apply rate limiters on a per-VM basis to share the bandwidth among them. 

\normalsize
\begin{table}[!t]
	\caption{Flow attributes and variables tracked in our mechanism}
	\centering
	\resizebox{9cm}{!}{
		\begin{tabular}{|c|c|}
		\hline
		Entry name (VM-to-VM) 	& Description \\\hline
		$source$ 			& IP address of source VM \\\hline
		$dest$ 			& IP address of destination VM\\\hline
		$out\_packet\_count$ 		& Sent packets count	\\\hline
		$ipr\_packet\_count$ 	& Received packets with ``IPR-bit'' mark\\\hline
		$ecn\_packet\_count$ 		& Received packets with ECN mark\\\hline\hline
		Variable name (per VM) 	& Description \\\hline
		$rate$ 	& The share rate or speed of NIC\\\hline
		$bucket$ 	& The capacity of the token bucket in bytes\\\hline
		$tokens$ 		& The number of available tokens to be used for transmission\\\hline
		\end{tabular}
		}
	\label{tab:flowandvariables}
	\end{table}
\normalsize 

HyGenICC deploys a flow table (for congestion control purpose) to track state information shown in Table \ref{tab:flowandvariables} on a VM-to-VM granularity (i.e., source VM-destination VM pairs). In addition, per-VM token-bucket state is used to enforce the VM's share of bandwidth. 

\subsection{VM detection and bandwidth allocation}
As soon as a VM's port becomes active (sending or receiving traffic), an associated entry is created in the flow table. Whenever a new VM becomes active on a given NIC, the NIC's nominal capacity is redistributed among the token buckets of active VMs to account for the new one. This is done by readjusting the rate and bucket size of all active VMs' token buckets on that NIC. Any extra traffic sent by the VM in excess of its share is simply dropped and resent later by the transport layer or otherwise a per-VM queue is used for holding the traffic for later transmission whenever the tokens are regenerated\footnote{We have experimented with both approaches and the queuing mechanism achieves slightly better performance which did not motivate its usage due to management and memory overhead.}. %

\subsection{Congestion Control Mechanism}

\begin{figure}[!t]
	\centering	 
		\includegraphics[scale=0.99, width=\columnwidth]{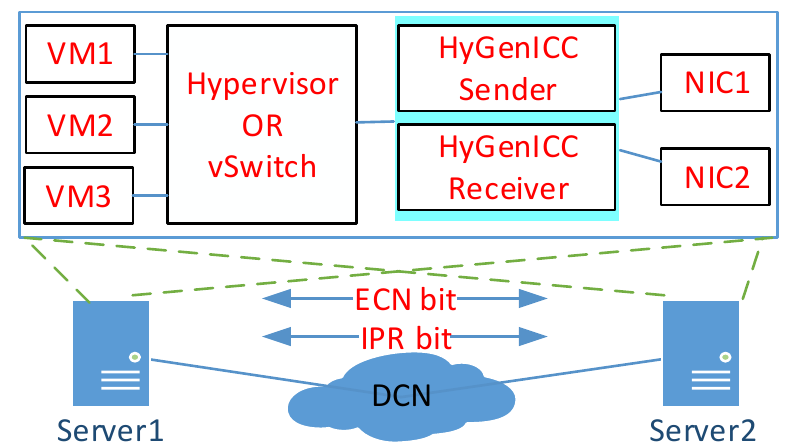}
	\caption{HyGenICC high-level system design}
	\label{fig:endhost-network}
\end{figure}

In practice, congestion may always happen within the network as shown in Figure~\ref{fig:endhost-network}, if the network is over-subscribed or does not provide full bisection bandwidth. HyGenICC therefore relies on readily available features in switches hardware\footnote{Most current commodity switches used in data centers are equipped with QoS mechanisms like Strict Priority (SP), Weighted Fair Queuing (WFQ) and Weighted Random Early Detection (WRED) in addition to the ability of ECN marking of IP packets.} %
, to convey congestion signals to the sources. To be more abstract, HyGenICC treats the datacenter network as a black box in which source servers inject traffic and the black box generates ECN marks in response to congestion towards the receivers. ECN marks are a fast proactive mechanism that can help in quickly detecting any congestion from a shared queue when buffers exceed a pre-configured queue occupancy threshold along a packet's path.  

HyGenICC uses the flow table to track, for each source-destination pair, the number of IP packets received with congestion notification marks, regardless of the type of transport protocol (TCP, UDP, or otherwise). This information is a valuable indication of the level of congestion along the path between the source VM and the destination VM starting at that particular NIC. Since HyGenICC implements a network-layer congestion control, any ECN or other marking used to track congestion is cleared before delivering the datagrams to the VM\footnote{Supposedly, If the tenants are willing to deploy ECN in their overlay networks then our mechanism should not clear ECN mark and let the transport layer handle it as well. We believe the two would not conflict rather complement each other as shown in the simulations.}. In addition, to force universal ECN marking along the path, all outgoing packets are marked with the ECN-enabled bit. HyGenICC typically creates a network layer congestion control loop between hypervisors and is fully transparent  to the overlying VM transport protocol.

At the receiver side, upon receiving ECN marks, HyGenICC needs to reflect the information back to the source to trigger reduction of the sending rate of that particular source VM. To avoid introducing any additional overhead and hinder the operation of any on-path middle-boxes by introducing a new protocol, we propose to piggyback the information on any returning data. For this we identify three types of traffic flows: TCP, which is by default bidirectional, other non-TCP bidirectional traffic and finally unidirectional traffic; for the three categories of traffic, we propose to use the unused reserved bit in the IP header ``IPR-bit'' of any reverse packet to reflect the ECN marking synchronously to the origin. While this might be sufficient for the first two categories of traffic to carry all marking back to the source, for the third category, there might be a dramatic imbalance in the forward traffic and reverse traffic leading to some proportion of forwarded markings not being reflected back. As a solution HyGenICC crafts a special small IP packet with header only (20 bytes of IP and 14 for Ethernet headers) and piggybacks explicitly the number of remaining ECN marks on the identification field of this IP packet. The IP protocol field is destined to an unused number that has meaning only for HyGenICC.

At the sender, to match the current sending rate to the congestion level in the network, upon receiving ``IPR-bit'' marks or the special packet, the source decreases the  VM's current allocated rate in proportion to the amount of marks and gradually increases the rate when no congestion bits are received in a period.

\section{Implementation}
\label{sec:implement}
As explained above, HyGenICC needs two mechanisms: rate limiters at the source server and congestion controller that run from source to destination server. These mechanisms can either be implemented in software, or hardware or a combination of both as necessary. %
We simplified the design and concepts of HyGenICC so that the built system is able to maintain line rate performance at 1-10Gb/s while reacting quickly to deal with congestion within a datacenter's short RTT time scale.

\subsection{HyGenICC sender}
HyGenICC sender processing is described in Algorithm \ref{alg:HYGENICCSender}.  At the senders HyGenICC tracks the \textit{rate}, the number of \textsl{tokens}, the depth of the \textsl{bucket} and the fill-rate variables per-VM per-NIC where the per-VM rate limiters are implemented as counting token buckets that have a rate $R(i,j)$ each, a bucket capacity $B(i,j)$ each and number of tokens $T(i,j)$ each. In addition, the sender will also handle the received congestion signals from different destinations on a per-source basis.

\normalsize
\begin{algorithm}[!ht]
\SetKwProg{Fn}{Function}{}{}

\Fn{Packet\_Departure($P,i,j$)}	
{
			\text{look up flow entry $f$ in flow table}
			\text{$T(i,j)=T(i,j) + R(i,j) \times (now() - f.senttime)$}
			\text{$T(i,j)=MIN( B(i,j) , T(i,j))$}
			\If{$T(i,j) \geq Size(P)$}
			{
				$T(i,j) = T(i,j) - Size(P)$
				$f.senttime = now()$
				\text{Enable ECN Capable bits (ECT) in IP header}
			}
			\Else
			{
				\text{Drop the packet}
			}
}

\Fn{Packet\_Arrival($P,i,j$)}
{
		    \text{look up flow entry $f$ in flow table}
			\If{Packet is congestion feedback message}
			{
				$f.feedback = f.feedback + int(P.data)$
			    $f.rbdetected = true$
			    $f.feedbacktime = now()$
			    \text{Drop the packet} 
			}
			\Else
			{ 
				\If{Packet is ``IPR-bit'' marked}
				{
					$f.feedback = f.feedback + 1$
					$f.rbdetected = true$
					$f.feedbacktime = now()$
					\text{Clear the mark and forward to the VM}
				}
			}				
}
\Fn{Timer\_timeout}
{
      \ForAll{flow $f$ in $FlowTable$}
      {
				\If{$now() - f.senttime \geq 1 sec$}
				{
					$f.active=false$
					\text{Reset $f$ entry in Flow Table}
					\text{Redistribute NIC capacity among active flows}
				}
	  }			
      \ForAll{Active flow $f$ in $FlowTable$}
      {
			   \If{now() - f.feedbacktime$\geq$ Congestion\_Timeout}
			   {
					$f.rbdetected = false$
			   }				
			   \If{$f.rbdetected == false$}
			    {
					 $R(i,j) = R(i,j) + scale(NIC\_Cap)$
				}
				\Else
				{ 
					\If{$f.feedback \geq 0$}
					{
						$R(i,j) = R(i,j)~-~(f.feedback~\times~scale(NIC\_Cap))$
					} 
				}
				\Else
				{
						$R(i,j) = R(i,j) + scale(NIC\_Cap)$ 
						$f.feedback=0$
				}
				$R(i,j) = MAX(0, MIN(Capacity\_Share, R(i,j)))$
	  }
}
\normalsize
\caption{HyGenICC Sender Algorithm}
\label{alg:HYGENICCSender}
\end{algorithm}

\subsubsection{Rate Allocation}
Initially, the installed on-system NICs are probed and the values of their nominal data rate $R(i)$, bucket capacity $B(i)$ and tokens $T(i)$ are calculated correspondingly. Thereafter, when packets start flowing from each source VM, NIC capacities are redistributed and a new capacity share ``$Capacity\_Share$'' is calculated and used to update the entries for each active VM in the rate, tokens and bucket matrices and the VM is marked as currently active on all outgoing physical NICs. 

After a certain time of inactivity\footnote{Inactivity timeout is set to 1 sec in simulations.}, the bucket entries for a VM are reset and its allocation is reclaimed and redistributed among currently active VMs. As shown in Table~\ref{tab:flowandvariables}, flow-table entries are established immediately after arrival of the first packet using source-destination IP address\footnote{We track the state of the communicating VM pairs not individual flows plus we find that the operations involved do not add burden on the hypervisor/vswitch.}. First, on arrival or departure of each packet $P$, its outgoing port $j$ and incoming port $i$ are detected. The current value of available tokens $T(i,j)$ is retrieved and replenished based on the elapsed time since the last transmission. Then, using the new $T(i,j)$, the packet is allowed for transmission if  $T(i,j) \geq size(pkt)$, in this case the packet length is deducted from $T(i,j)$, otherwise the packet is dropped.

\subsubsection{Congestion Reaction}
The sender module reacts on regular intervals to incoming ``IPR-bit'' and cuts the sending rate in proportion to the amount of marking received. Hence, sources causing congestion in the network will receive ``IPR-bit'' signals and will react by decreasing their sending rates proportionally until the congestion subsides and congestion signals start disappearing at which time sources start to gradually increase their rates. The process will increase the rate conservatively, and if no feedback arrives within $Congestion\_Timeout$ seconds, the rate is increased fast until it reaches its ``Capacity\_Share'' or an ``IPR-bit'' is detected again. Function ''scale(NIC\_Cap)`` is used to scale the amount of rate increase and decrease to account for a single packets transmission over a single RTT (i.e., 1000 bytes over 1 Gb/s in average RTT range of 100$\mu$s-10ms would give us $\approx$ 80-8 Mb/s of increments).

\subsection{HyGenICC receiver}
At the receiver, HyGenICC needs to track incoming congestion ECN marks from the network on a per-source-destination basis and feed this information back by piggybacking it on outgoing packets heading back to corresponding sources. Hence, the operations of the receiver  is quite simple and does not incur much processing overhead onto incoming traffic. The receiver processing is described in Algorithm~\ref{alg:HYGENICCReciever}. 

\normalsize

\begin{algorithm}[!ht]
\SetKwProg{Fn}{Function}{}{}

\Fn{Packet\_Arrival($P,i,j$)}
{
			\text{look up flow entry $f$ in flow table}
			\If{Packet is ECN marked}
			{
				$f.ecnmarks = f.ecnmarks + 1$
				\text{Clear the mark and forward to the VM}
			}
			\If{now() - f.feedbacksenttime $\geq$ feedback\_timeout}
			{
				\text{Create IP feedback message and send to $f.source$}
				$f.feedbacksenttime = now()$
				$f.ecnmarks=0$
			}
}
\Fn{Packet\_Departure($P,i,j$)}
{	
			\text{look up flow entry $f$ in flow table}
			\If{$f.ecnmarks \geq 1$}
			{
				 \text{Set ``IPR-bit'' flag in IP header}
				 $f.feedbacksenttime = now()$
				 $f.ecnmarks = f.ecnmarks - 1$				
			}
}
\normalsize
\caption{HyGenICC Receiver Algorithm}
\label{alg:HYGENICCReciever}
\end{algorithm}

Each incoming packet is checked for ECN mark and the number of packets with and without the mark are traced in the flow table, Table~\ref{tab:flowandvariables}, and immediately the ECN mark is cleared before re-injecting the packet in the normal packet processing path. For each ECN marked packet, an IPR-bit mark is reflected in the first available outgoing packet to that destination (it could be a TCP ACK if the flow is TCP or a UDP reply data packet)  until all the ECN marks are cleared. However, when ingress and egress traffic are out of balance on a given flow, non-reflected ECN marks may start to accumulate at the receiver, to address this issue, we periodically use an explicit ICMP-like feedback packet to convey the remaining amount of ECN marks to the source. On a regular intervals close to an  RTT, we scan through the flow table asynchronously for any flow with remaining ECN marks and that has not sent any feedback for a period of $Feedback\_Timeout$. If any is found, then an IP packet is created with unused protocol ID value and the current value of ECN marks added as a 2-bytes payload of this packet addressed to the source of the flow. This event is infrequent and unlikely to exist but if so, will not incur much network overhead as the packet size would be 36 bytes (14-bytes Ethernet header + 20-bytes IP header + 2-bytes payload data). To compress further the explicit feedback, the 2 bytes payload can be piggybacked instead in the IP header identification field.

\section{Proposed Methodology}
\label{sec:method}

\begin{figure}[t]
	\centering	 
	\includegraphics[width=0.9\columnwidth]{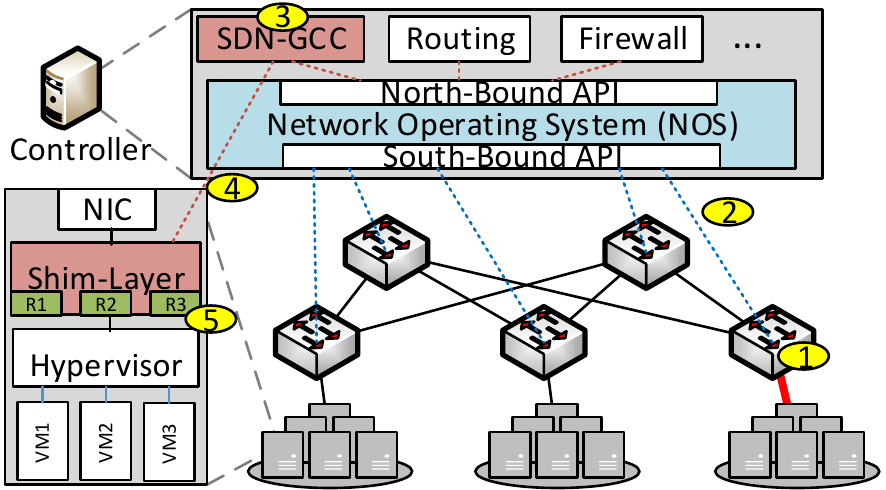}
	\caption{SDN-GCC high-level system design: 1) congestion point; 2) network statistics; 3) congestion tracking; 4) congestion notification; 5) rate adjustment.}
	\label{fig:SDN-GCC}
\end{figure}

\figurename~\ref{fig:SDN-GCC} shows SDN-GCC's system design which is broken down into two parts: a network application that runs on the SDN controller (network OS). It is responsible for monitoring network states by querying the switches periodically via SDN's standard southbound API and signaling congestion; and a hypervisor-based shim-layer, that is responsible of enforcing per-VM rate control in response to congestion notification by the control application. The following scenario sketches the SDN-GCC cycle:
\begin{inparaenum}[1)]
\item Whenever the total incoming load exceeds the link capacity, the link (in-red) becomes congested implying that senders are exceeding their allocated rates. 
\item SDN-switches sends to the network OS periodic keep-alive and statistics through the established control plane between them (e.g., OpenFlow or sFlow). Whenever necessary, the switch would report the amount of congestion experienced by each output queue of its ports.
\item The SDN-GCC application co-located with the network OS (or alternatively communicating via the north-bound API) tracks congestion events in the network.
\item SDN-GCC application communicates with the SDN-GCC shim-layer of the sending servers whose VMs are causing the congestion.
\item SDN-GCC shim-layer takes a corrective action by adjusting the rate-limiter of the target VM.
\end{inparaenum}

We start from a single end-host (hypervisor) connecting all VMs where bandwidth contention happens at the output link (i.e., when multiple senders compete to send through the same output NIC of the virtual switch). The hypervisor needs to distribute the available NIC's capacity among VMs and ensure compliance of the VMs' weights with the allocated shares. Hence it employs a mechanism to apply rate limiters on a per-VM basis. Table \ref{tab:variables} shows the variables needed to implement a per-VM token-bucket rate limiter. Ideally, when a virtual port becomes active, its variables are initialized and the NIC's nominal capacity is redistributed among the rate limiters of currently active VMs by readjusting the rate and bucket size of all active VMs' token buckets on that NIC. Then we need to extend the allocation of single hypervisor to account for the in-network congestion caused by a network of hypervisors managing tenants' VMs.

\begin{table}[!ht]
	\caption{Variables and parameters used by the SDN controller application and end-host shim-layer of SDN-GCC}
	\centering
	\resizebox{8.7cm}{!}{
		\begin{tabular}{|c|c|}
		\hline
		\textbf{Parameter name} 	& \textbf{Description} \\\hline
		$T_o$ 		& Timeout for Flow inactivity period\\\hline
		$T_c$ 		& Timeout for Congestion grace period \\\hline  
		$T_i$		& Timeout of Congestion monitor period \\\hline   
		\hline		
		\textbf{Variable name (Shim-Layer)} 	& \textbf{Description} \\\hline
		$source$ 			& IP address of source VM \\\hline
		$vport$				& virtual port connecting VM \\\hline
		$rate$ 	& The allocated sending rate\\\hline
		$bucket$ 	& The capacity of the token bucket in bytes\\\hline
		$tokens$ 		& The number of available tokens\\\hline
		$sent\_time$ 		& The time-stamp of last transmission\\\hline
		\hline
		\textbf{Variable name (SDN APP)} & \textbf{Description} \\\hline
		$SWITCH$ 		& List of the controlled SDN switches \\\hline
		$SWITCHPORT$ 		& List of the ports on the switches \\\hline
		$DSTSRC$ 		& List of destinations to sources pairs \\\hline
		$IPTOPORT$ 		& List of IP to switch port pairs \\\hline
		$MARKS$ & ECN marks reading of for each switch port \\\hline
		\end{tabular}
		}
\label{tab:variables}
\end{table}

\normalsize

In practice, congestion may always happen within the data center network, if the network is over-subscribed or does not provide full bisection bandwidth. SDN-GCC in an effort to account for this limitation, relies on readily available functionality in SDN switches to convey congestion events to the controller. To elaborate more, SDN-GCC controller can keep a centralized record of congestion statistics by periodically collecting state information from the switches as shown in Table~\ref{tab:variables}. ECN marking is chosen as a fast live-congestion indication to signal the onset of possible congestion at any shared queue. However, Usage of RED and ECN marking could be avoided if drop-tail AQM keeps statistics of backlog exceeding a certain pre-set threshold.  

SDN-GCC application running on top of the network OS, keeps record of each network-wide state information (e.g., congestion points). Hence, it can infer the bottleneck queues based on this information and make intelligent decisions accordingly. Whenever necessary, it sends special congestion notifications to the shim-layer to adjust the sending rate of the affected VM. Upon receiving any congestion notification The shim-layer reacts by adjusting VM's rate-limiter proportionally to the congestion level in the network and gradually increases the rate when no more congestion messages are received.

\section{Design and Implementation}
\label{sec:implement}

As explained above, SDN-GCC needs two components: shim-layer at the servers and the control application that runs on top of the network OS. These mechanisms can either be implemented in software, or hardware or a combination of both as necessary. We simplified the design and concepts of SDN-GCC so that the built system is able to maintain line rate performance at 1-10Gb/s while reacting quickly to deal with congestion within a reasonable time.

\begin{algorithm}[t]
\caption{SDN-GCC Shim-layer}
\label{alg:SDGENCCSender}

\SetKwProg{Fn}{Function}{}{}

\Fn{$Normal\_Packet\_Arrival(P,src,dst)$}	
{
	\tcc{i is NIC and j is VNIC index}
		$T(i,j)=T(i,j) + R(i,j) \times (now() - f.sent\_time)$\;
		$T(i,j)=MIN(B(i,j) , T(i,j))$\;
		\text{Enable ECN-capable bits (ECT) in IP header}\;
		\If{$T(i,j) \geq Size(P)$}
		{			
				$T(i,j) = T(i,j) - Size(P)$\;
				$sent\_time(i,j) = now()$\;
		}
		\Else
		{
			\text{Queue until token regeneration OR Drop}\;		
		}	
}
\Fn{$Control\_Packet\_Arrival(P,i,j)$}
{
	 \If{\text{Packet has congestion notification message}}
	  {
	  		$marks = int(msg)$\;
			\If{$marks \geq 0$}						
			{
				$cong\_detected(i,j) = true$\;
				$elapsed\_time = now() - cong\_time(i,j)$\;
				$mark\_rate = \frac{marks}{elpased\_time}$\;
				$R(i,j) = R(i,j)-(mark\_rate \times scale(C))$\;
				$R(i,j) = Max(R_{min}, R(i,j))$\;
				$cong\_time(i,j) = now()$\;
			}			
	}
	\Else
	{
		\text{Send to normal packet processing}\;
	}
}	

\Fn{$State\_Update\_Timeout()$}
{
     \ForAll{i in NICs and j in VNICs}
     {
      	\eIf{$now() - cong\_time(i,j) \geq  T_c$}
		{
			$cong\_detected(i,j) = false$\;
		}		
		{
			\If{$now() - sent\_time(i,j) \geq T_o$}
			{
			  $active(i,j) = false$\;
			  \text{redistribute NIC capacity among active flows}\;
			}
		}			
		\If{$active(i,j)~\&\&~cong\_detected(i,j)$}
		{
					$R(i,j) = R(i,j) + scale(C)$\;
					$R(i,j) = MIN(E(i), R(i,j))$\;	
		}				
	}
}
\end{algorithm}

\subsection{SDN-GCC End-Host Shim-Layer}
SDN-GCC shim-layer processing is described in Algorithm \ref{alg:SDGENCCSender}. The major variables it tracks are the \textit{rate}, the number of \textit{tokens} and the depth of the \textit{bucket} variables per-VM per-NIC where the per-VM rate limiters are implemented as counting token buckets where virtual NIC $j$ has a rate $R(i,j)$, bucket depth $B(i,j)$ and number of tokens $T(i,j)$ on physical NIC $i$. In addition, the shim-layer will also translate the received congestion message from the controller on a per-source basis.

Initially, the installed on-system NICs are probed and the values of their nominal data rate $R(i)$, and bucket size $B(i)$ are calculated. Thereafter, when the first packet is intercepted from a new VM, NIC capacity is redistributed and a equal-share of capacity ``$E(i)$'' is calculated. The new value $E(i)$ is used to re-distribute the allocated rate for each active VM and then the new VM is marked as active\footnote{Typically, after a certain time of inactivity (e.g., 1 sec in our simulation), the variables used for VM tracking are reset and the rate allocations are redistributed among currently active VMs.} As shown in Table~\ref{tab:variables}, the state of the communicating VM is tracked only through token bucket and congestion specific variables. The shim-layer algorithm shown in Algorithm~\ref{alg:SDGENCCSender} is located at the forwarding stage of the stack, on arrival or departure of a packet $P$, it detects the packet's outgoing port $j$ and incoming port $i$. Before $P$'s departure, the available tokens $T(i,j)$ is refreshed based on the elapsed time since the last transmission. The packet is then cleared for transmission if $T(i,j) \geq size(P)$, in which case $size(P)$ is deducted from $T(i,j)$. Otherwise, $P$ is simply dropped\footnote{Packets could be queued for later transmission, however, this approach adds  a large overhead on the end-hosts}. The shim-layer intercepts only the special congestion message.

For each incoming notification, the algorithm cuts the sending rate in proportion to the rate of marking received (capped by $R_{min}$) which is a parameter set by the operator. Typically, this values is chosen with respect to the minimal bandwidth guarantee provided by the operator. Hence, as sources cause more congestion in the network, the amount of marks received increases and as a result the sending rates of such sources decrease proportionally until the congestion subsides. When congestion messages become less frequent, or after a pre-determined timer $T_c$ elapses, the algorithm starts to gradually increase the VMs' source rate conservatively. The rate is increased until it reaches $E(i)$, or congestion is perceived again, leading to another reduction cycle. Function ''scale(C)" is used to scale the amount of rate increase and decrease proportional to the NICs rate and to smooth out large variations in rate dynamics. %

\subsection{SDN Network Application}
SDN-GCC relies on a SDN network application to probe for congestion statistics on a regular basis from the queues of the SDN switches in the network. The application sends notification messages towards the VMs 
that are causing congestion on a given port. This is accomplished by building a special message with those particular VMs as destinations with the data indicating the amount of marking they have caused. These messages are never delivered to the VMs and are actually intercepted by the shim-layer in the hypervisor. For simplicity, we assume that each of the involved VMs contribute equally to the congestion, and hence, the marks are divided equally among source VMs. %
SDN-GCC Controller shown in Algorithm~\ref{alg:SDGenCCController} is an event-driven mechanism that handles two major events: packet arrivals of unidentified flows (miss-entries) from switches and congestion monitoring timer expiry to trigger warning messages to the involved sources if necessary.  
\begin{enumerate}
	\item \textbf{Upon a packet arrival:} extract the necessary information to establish source to destination $DST\_SRC$ relationship and destination to port relationship $IPTOPORT$. This is necessary to establish associations between congested ports and corresponding sources. In addition, The timer for congestion monitoring is re-armed if it was not already. 
	\item \textbf{Congestion monitoring timer expiry:} for each switch $sw$, the controller probes for marking statistics through OpenFlow or sFlow protocols by calling function $read\_marks(sw)$. The new marking rate of each switch port $p$ is calculated~\cite{OpenFlow1.5}. For each port, if there are new markings (due to congestion), then the controller needs to advertise this to all related sources. Thus we first retrieve the destination list of this port via function $get\_all\_dst(sw,p)$ and then for each destination retrieve the sources  using  $get\_all\_src(dst)$. The controller now piggybacks on any outgoing control message or crafts an IP message consisting of an Ethernet Header (14 bytes), an IP header (20 bytes), and a payload (2-byte) containing the number of ECN marks that have been observed in the last measurement period, divided by the number of sources. This message is created for each source concerned (sending through the port $p$ experiencing congestion) and sent with the source IP of the dest. VM and dest. IP of the source VM (which allows the shim-layer to identify the appropriate forwarding ports of source VM). %
\end{enumerate}

\begin{algorithm}[t]
\caption{SDN-GCC SDN application}
\label{alg:SDGenCCController}
\small
\SetKwProg{Fn}{Function}{}{}

\Fn{$Packet\_Arrival(P,src,dst)$}
{			
	\If{P is an IP packet of a new source-destination pair}
	{
		$DSTSRC[P.src] = P.dst$\;
		$IPTOPORT[P.src] = P.in\_port$\;
		\lIf{\text{Timer is not active}}
		{
			\text{start} $CM\_Timer(T_i)$\
		}
	}			
}
\Fn{$Congestion\_Monitor\_Timeout()$}
{
	\ForAll{$sw~in~SWITCH$}
	{
		$sw\_marks=read\_marks(sw)$\;
		\ForAll {$p~in~SWITCH\_PORT$}
		{
			$\alpha=MARKS[sw][p]~-~sw\_marks[p]$\;
			$MARKS[sw][p]=sw\_marks[p]$\; %
			\If{$\alpha~>~0$}
			{
				$DST\_LIST=get\_all\_dst(sw,p)$	\;
				\ForAll {$dst~in~DST\_LIST$}
				{
					$SRC\_LIST[dst]=get\_all\_src(dst)$\;
					$\beta=\beta~+~size(SRC\_LIST[dst])$\;
				}
				\If{$\beta~>~0$}
				{
					$m=\frac{\alpha}{\beta}$\;
					\ForAll {$dst~in~DST\_LIST$}
					{
							\ForAll {$src~in~SRC\_LIST[dst]$}
							{
								$msg=MSG~(~m~,~dst~,~src~)$\;
								$send~msg~to~src$\;
							}
					}
				}
			}
		}
	}
	\text{Restart} $CM\_Timer(T_i)$\;
}
\end{algorithm}

\subsection{Implementation and Practical Issues}
Any traffic sent by the VM in excess of its share can either be queued or simply dropped and resent later by the transport/application layer. In the former case, an extra per-VM queue is used for holding the traffic for later transmission whenever the tokens are regenerated. We tested both approaches and the queuing mechanism turned out to achieve marginally better performance which, in view of the complexity it adds, did not motivate its need. If ECN marking is not in use end-to-end, all outgoing data packets must be marked with the ECT bit. In addition, the shim-layer needs to clear any ECN marking used to track congestion before delivering the packets to the target VMs. %

SDN-GCC is distributed between the network application and the shim-layer with very low computational complexity and can be integrated easily in any network whose infrastructure is based on SDN. Due to recent advancement of memory speeds, the throughput of internal forwarding (e.g., Open vSwitch (OvS)) of commercial desktop/server is 50-100 Gb/s, which is fast enough to handle 10's of concurrent VMs sharing a single or few physical links. Hence, the overhead of the shim-layer functions added to the OvS would not overload the CPU. In addition, the shim layer at the hypervisor require operations of $O(1)$ per packet, as a result the additional overhead is insignificant. The network application is also of low complexity making it ideal for fast response to congestion (within few milliseconds time scale). 

In multi-path routing, the SDN application with global view can evaluate congestion on a path-by-path basis. Consequently, the shim-layer can adapt rates to each path which it can identify via the 5-tuple hash of the packets. Finally, the control plane communications in SDN networks is typically out-of-band i.e., different from the data plane \cite{Panda2013}, hence fast reaction to congestion is possible and notification messages are not interrupted by in-network middle-boxes.

In data centers, the internet-facing incoming/outgoing connections are typically isolated from intra-data center connections via connection splitting at the front-end or proxy servers. Hence, the intra-data center congestion is primarily attributed to flows running in the data center

\section{Simulation Analysis}

\label{sec:evaluate}
In this section, we study the performance of the proposed scheme via ns2 simulation in network scenarios with a high bandwidth-low delay. We examine the performance of a tagged VM that uses New-Reno TCP with SACK-enabled. The tagged TCP connection  competes with other VMs running similar New-Reno TCP, DCTCP, or UDP in four cases:
\begin{inparaenum}[\itshape 1) \upshape]
\item a setup that uses RED AQM with non-ECN enabled TCP; 
\item a setup that uses RED AQM with ECN enabled TCP;
\item a setup that uses HyGenICC as the traffic control mechanism; and
\item a setup that uses proposed SDN-GCC framework. 
\end{inparaenum}
For HyGenICC, there is a single parameter settings of timeout interval for updating flow rates which should be larger than a single RTT, in the simulation this value is set to 500 $\mu$s (i.e., 5 RTTs). However, in the case of SDN-GCC, timeout interval for congestion monitoring and reporting application of the controller is set to a large value of 5ms (i.e., 50 RTTs). We set the minimum rate $R_{min}$ to 1\% of the share of the sender VM on the underling NIC. In all simulation experiments, we adjust RED parameters to perform ECN marking based on the instantaneous queue length at the threshold of 20\% of the buffer size.

\subsection{Simulation Setup}
We use ns2 version 2.35 \cite{NS2} which we patched to include the different mechanisms. We use in all our simulation experiments link speed of 1 Gb/s for stations, small RTT of 100 $\mu$s and the default TCP $RTO_{min}$ of 200 ms. 
We use a single-rooted tree (Dumbbell) topology with single bottleneck link of 1Gb/s at the destination, and run the experiments for a period of 30 sec. The buffer size of the bottleneck link is set to be more than the bandwidth-delay product in all cases (100 Packets), the IP data packet size is 1500 bytes.

\subsection{Simulation Results and Discussion}

For clarity, we first consider a toy scenario with 4 elephant flows (the tagged flow and 3 competitors). In the experiments, the tagged FTP flow uses TCP NewReno and competes either with 3 FTP flows using TCP NewReno, DCTCP or 3 CBR flows using UDP. Competitors start and finish at the $0^{th}$ and $20^{th}$sec respectively, while the tagged flow starts at the $10^{th}$sec and continues until the end. Hence, from 0 to 10s (period 1) only the competitors occupy the bandwidth, from 10s to 20s (period 2) the bandwidth is shared by all flows, and from 20s to 30s (period 3) the tagged flow uses the whole bandwidth. This experiment demonstrates work conservation, sharing fairness, and convergence speed of SDN-GCC compared to other setups. 		

\begin{figure}[t]
\captionsetup[subfigure]{justification=centering}
\centering
	\begin{subfigure}[ht]{0.8\columnwidth}
                \includegraphics[width=\textwidth, height=5cm]{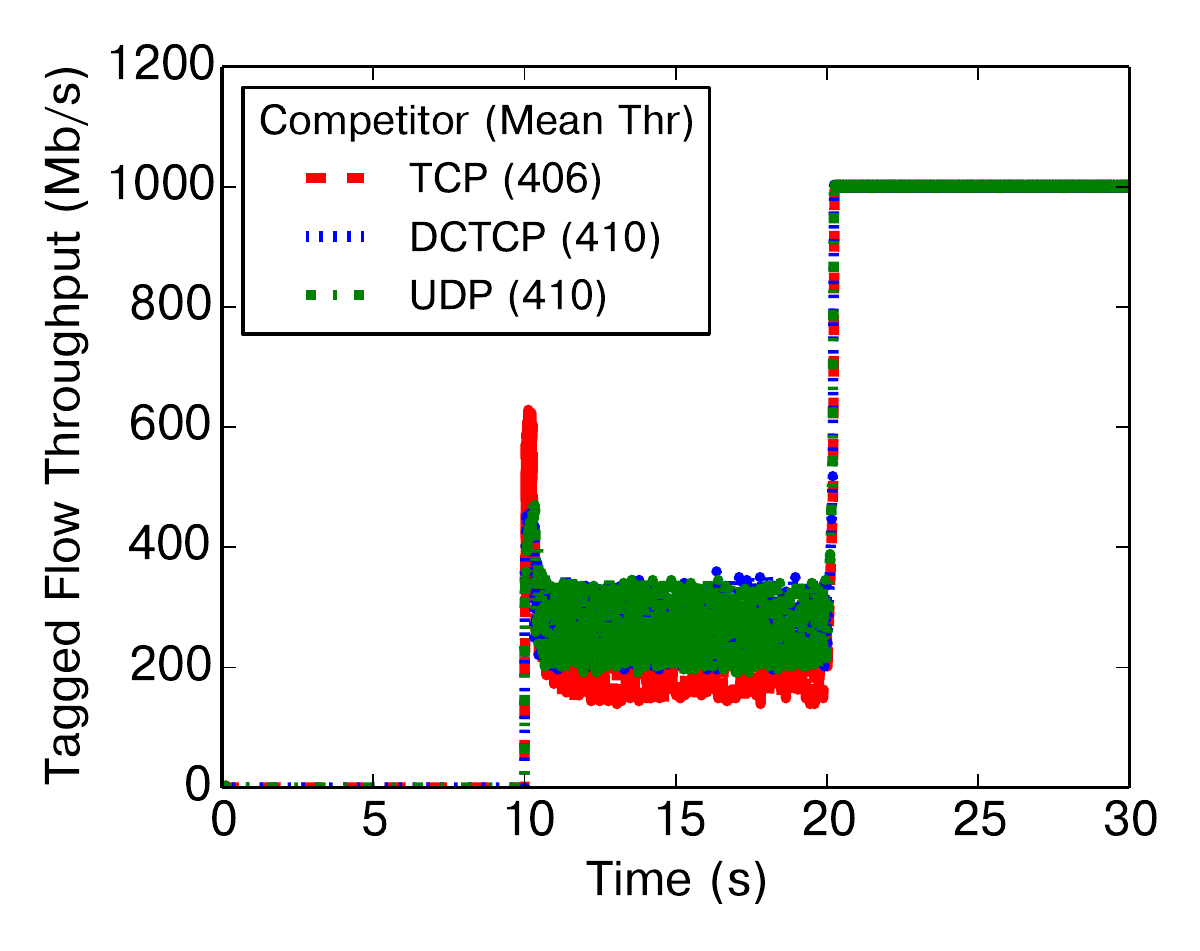}
                \caption{HyGenICC enabled network}
                \label{fig:HyGenICC}						
        \end{subfigure}
			  \\
        \begin{subfigure}[ht]{0.8\columnwidth}
			\includegraphics[width=\textwidth, height=5cm]{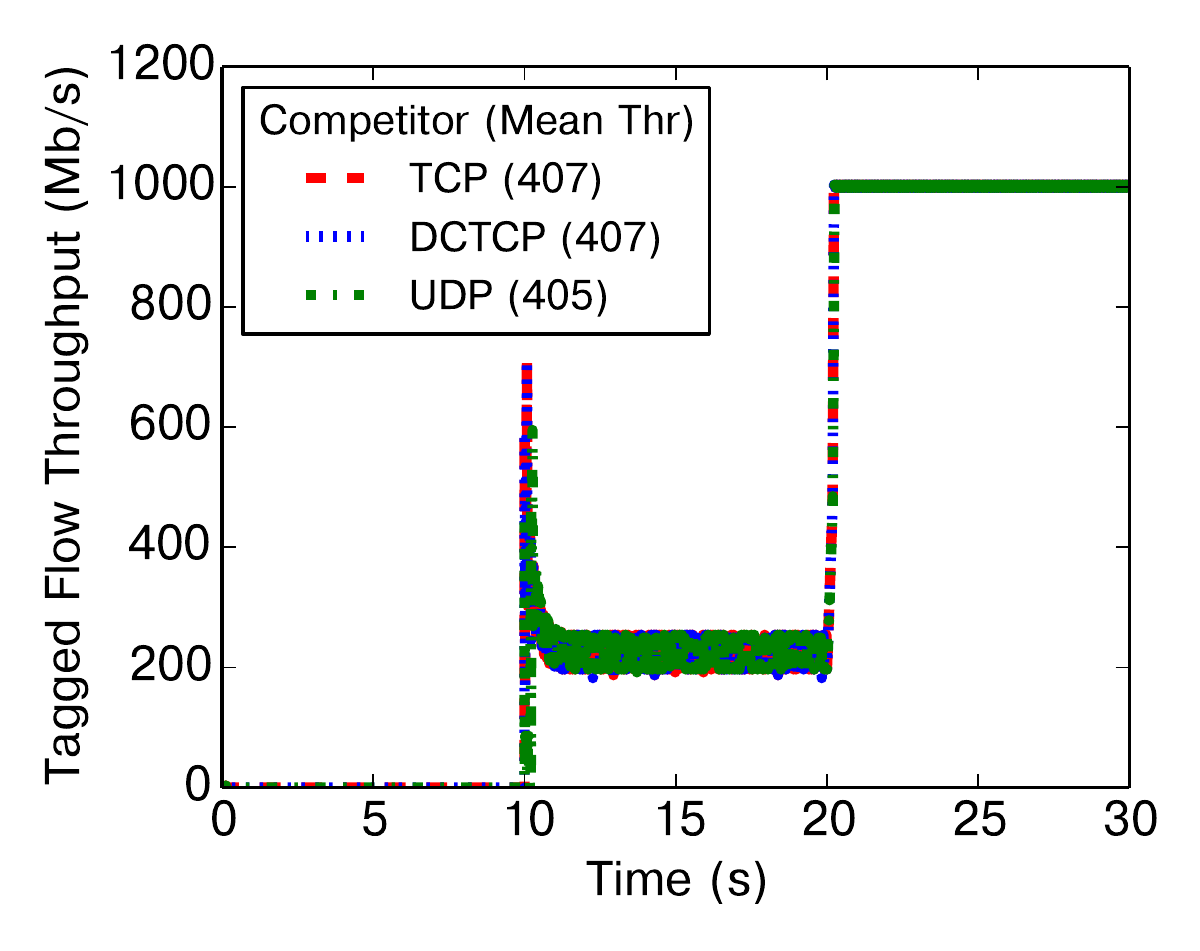}
                \caption{SDN-GCC enabled network}
                \label{fig:SDGenCC}
        \end{subfigure}
				\caption{The instantaneous and mean goodput of the tagged TCP flow while competing with 3 senders that use either TCP, DCTCP or UDP. In interval [0,10] the competitors are active, in [10,20] all flows are active and in [20,30] only the tagged TCP flow is active.}
				\label{fig:4flows}
\end{figure}

\begin{figure*}[ht]
\captionsetup[subfigure]{justification=centering}
\centering
        \begin{subfigure}[ht]{0.32\textwidth}
                \includegraphics[width=\textwidth, height=4.5cm]{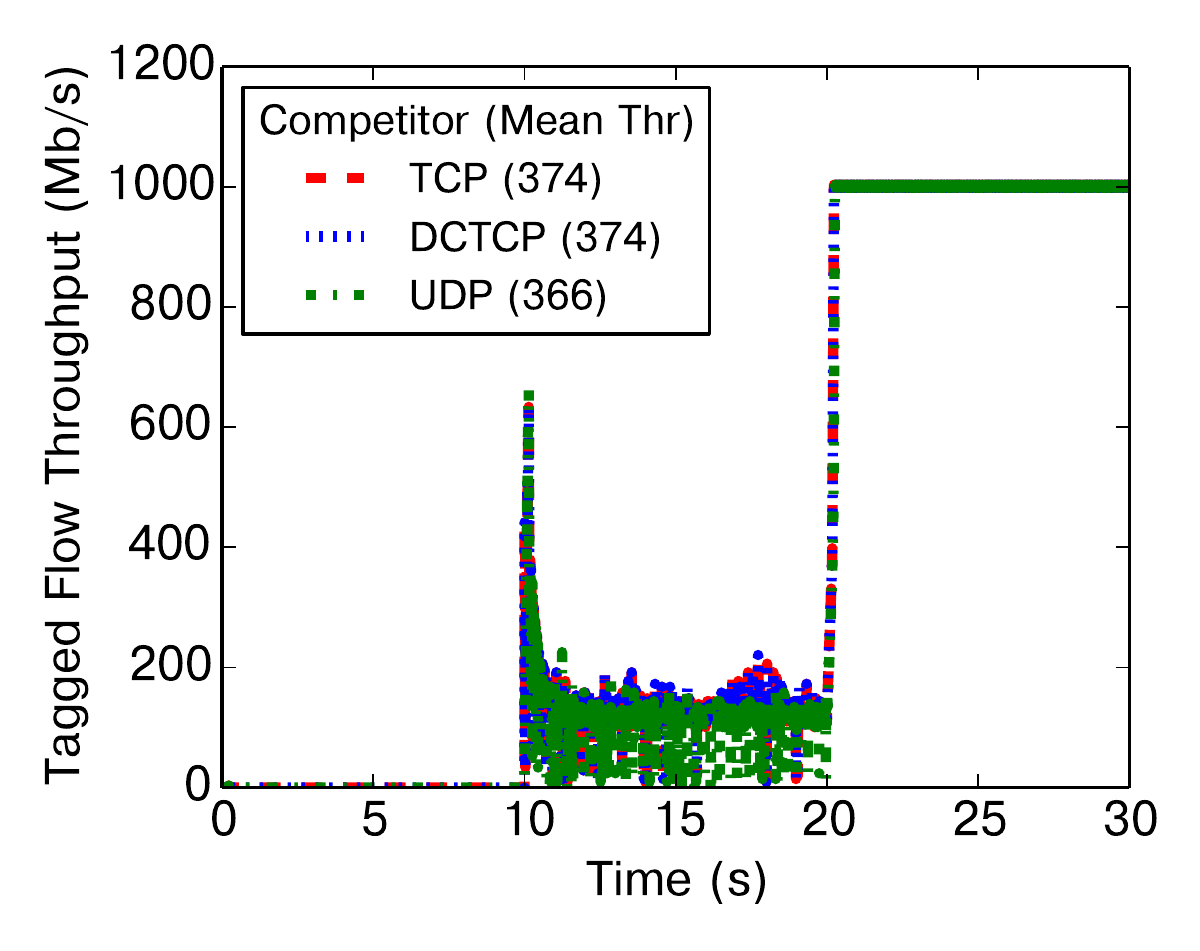}
                \caption{8 senders scenario}
                \label{fig:8sender}							
        \end{subfigure}
			  \hfill
        \begin{subfigure}[ht]{0.32\textwidth}
		\includegraphics[width=\textwidth]{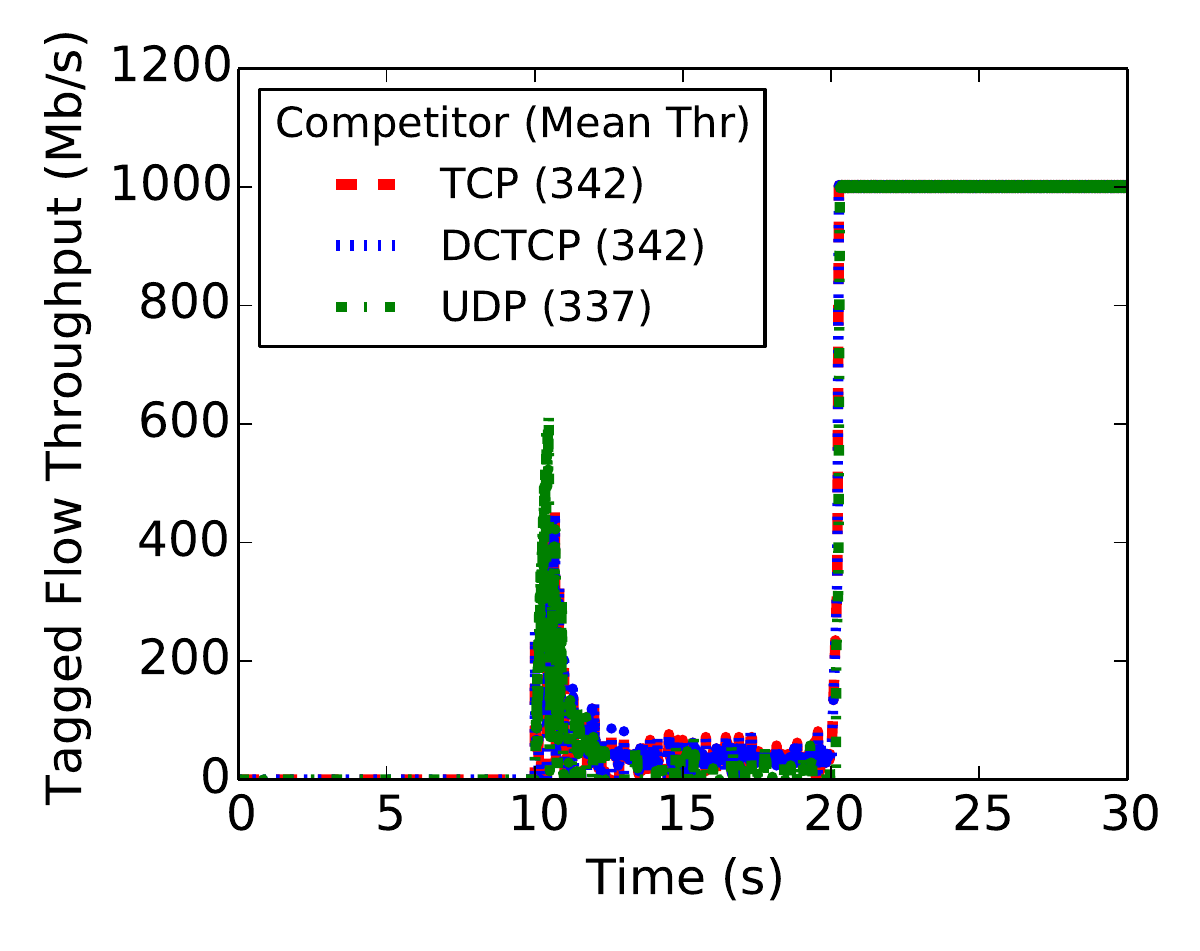}
                \caption{16 senders scenario}
                \label{fig:16sender}
        \end{subfigure}
        \hfill
        \begin{subfigure}[ht]{0.32\textwidth}
		\includegraphics[width=\textwidth]{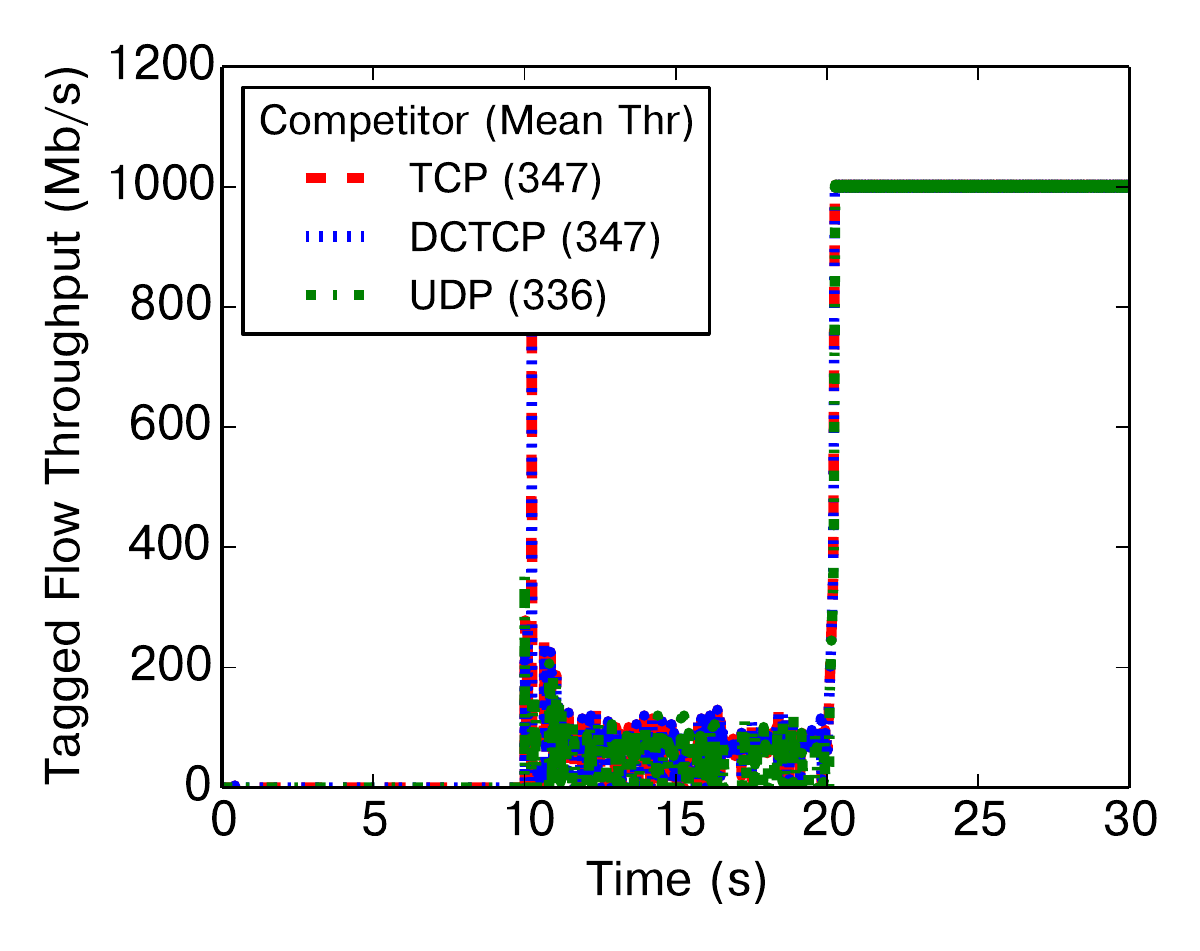}
                \caption{32 senders scenario}
                \label{fig:32sender}
        \end{subfigure}
	\caption{The instantaneous and mean goodput of the tagged TCP flow competing with 7, 15 and 31 senders using TCP, DCTCP or UDP}
	\label{fig:mult-sender}
\end{figure*}
						
\figurename~\ref{fig:4flows} shows the instantaneous goodput of the tagged TCP flow along with the mean goodput with respect to its competitor (in the legend, the optimal average goodput of tagged TCP would be 0Mb/s for period 1, 250Mb/s for period 2, 1000Mb/s for period 3 and 416Mb/s for all the periods). As shown in \figurename~\ref{fig:noecn}, without any explicit rate allocation mechanisms and without ECN ability, TCP struggles to grab any bandwidth when competing with DCTCP and UDP flows as DCTCP and UDP are more aggressive. \figurename~\ref{fig:ecn} suggests that ECN can partially ease the problem, however the achieved throughput reaches the allocated share only when the competitor uses the same TCP protocol. This can be attributed to the fact that TCP reacts conservatively to ECN marks unlike DCTPC which reacts proportionally to the fraction of ECN marks. Simulation with a static rate limit of 250 Mb/s (fair-share), show that a central rate allocator assigning rates per VM can achieve perfect rate allocation with no work-conservation (Utilization is 250 Mb/s in period 3). \figurename~\ref{fig:HyGenICC} shows that HyGenICC thanks to its distributed and live adaptive rate limiters, can respond effectively to congestion events. Finally, \figurename~\ref{fig:SDGenCC} suggest a similar result as HyGenICC can be achieved with the help of a regular control messaging from a central controller whenever necessary. Hence, SDN-GCC can efficiently leverage its global view of network status to dynamically adjust the rate limiters controlling the competing flows that cause congestion and yet achieve work conservative high network utilization.

SDN based schemes are questioned for their scalability which is currently under active research \cite{Karakus2017}. We conduct the previous experiment however we increase the number of competitors to 7, 15 and 31.  \figurename~\ref{fig:mult-sender}  suggests that SDN-GCC can scale well with an increasing number of senders. The tagged TCP flow and competing flows, starting at $10^{th}$, adjust their rates due to the incoming control messages when the controller starts observing congestion in the network. The adjustment messages trigger flow rate changes up and down until they reach the equilibrium point where sources start oscillating slightly around the target share of $\approx\frac{1Gb}{8} \approx 125Mb$, $\approx\frac{1Gb}{16} \approx 62.5Mb$ and $\approx\frac{1Gb}{32} \approx 31.25Mb$ for the 8, 16 and 32 sources respectively. 

\begin{figure*}[ht]
\captionsetup[subfigure]{justification=centering}
\centering
		\begin{subfigure}[ht]{0.32\textwidth}
                \includegraphics[width=\textwidth]{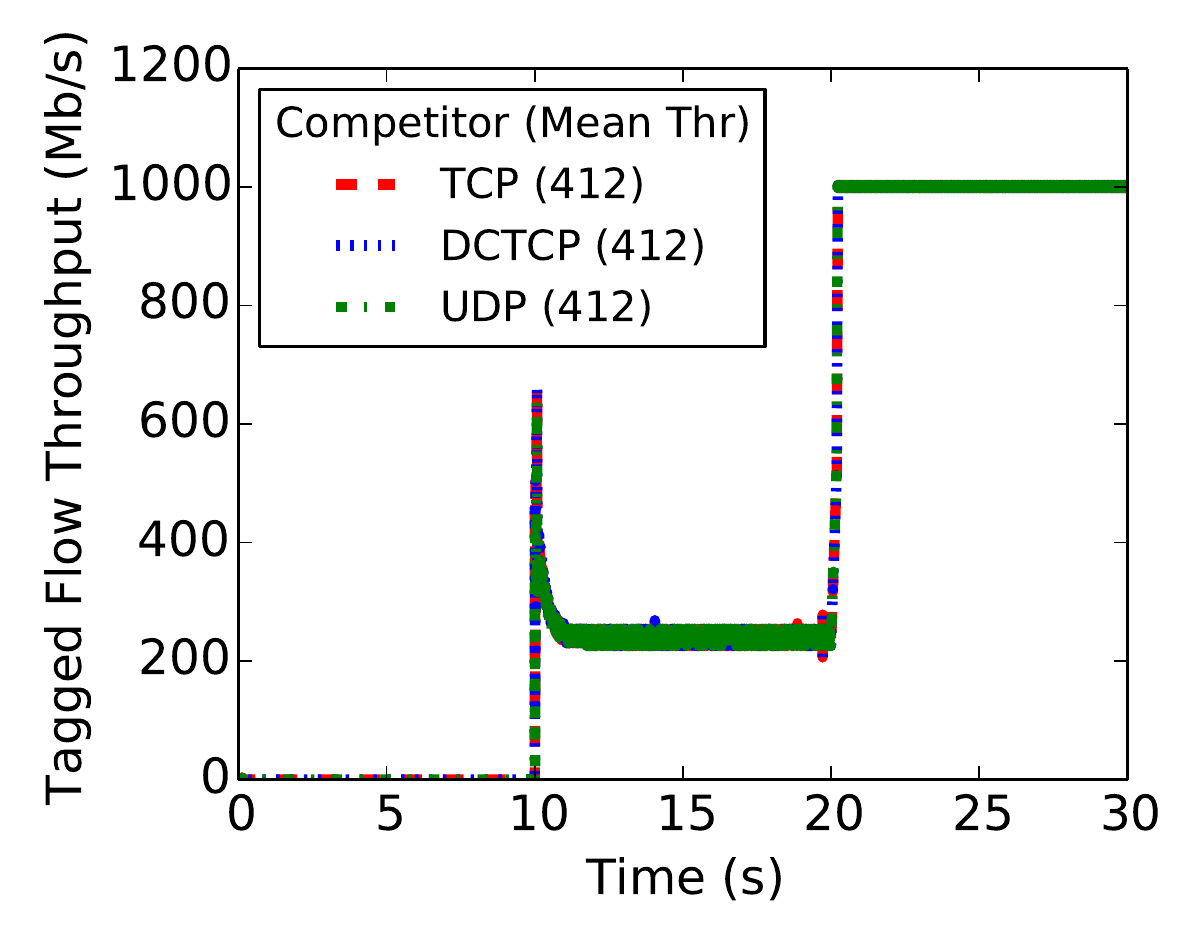}
                \caption{1ms monitor interval}
                \label{fig:1ms}							
        \end{subfigure}
        \hfill
        \begin{subfigure}[ht]{0.32\textwidth}
                \includegraphics[width=\textwidth]{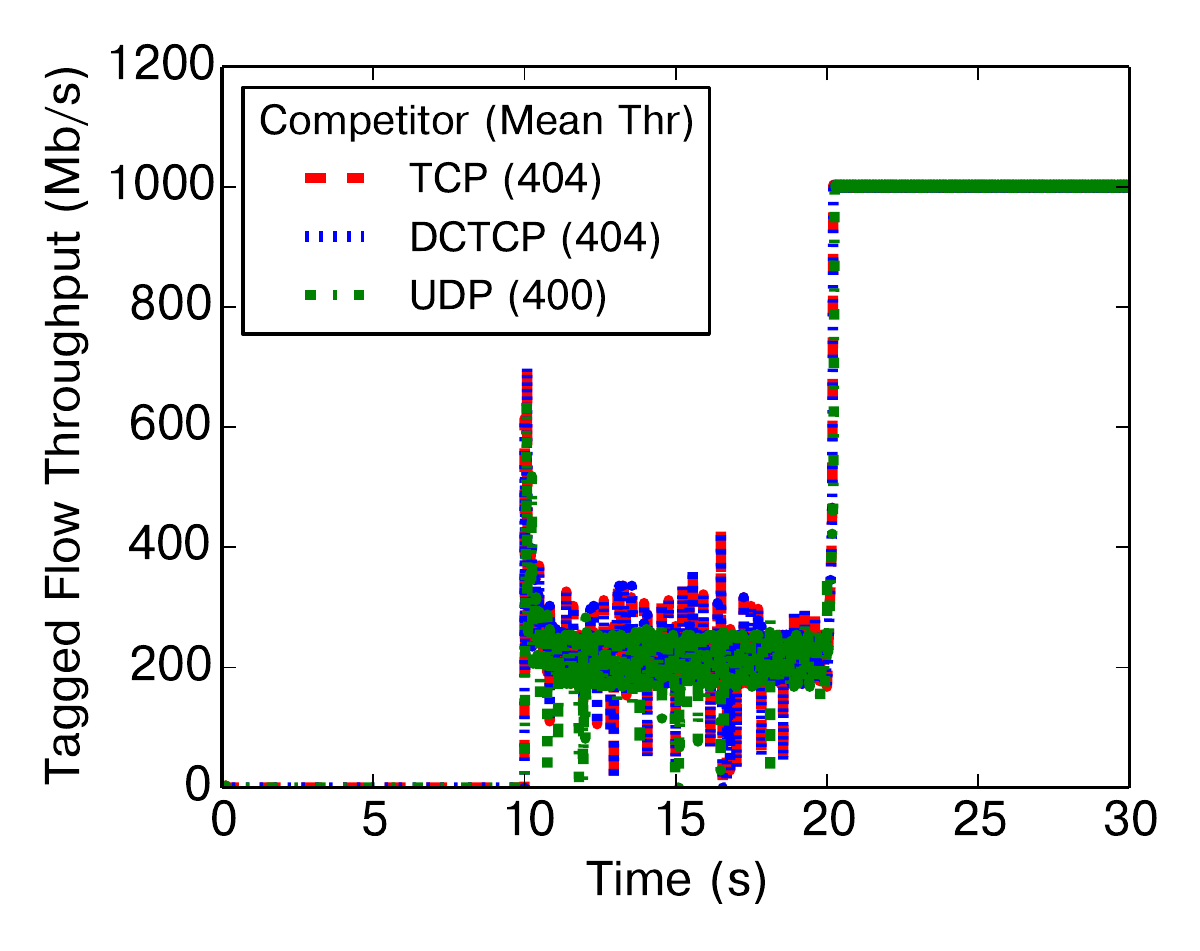}
                \caption{10ms monitor interval}
                \label{fig:10ms}							
        \end{subfigure}
			  \hfill
        \begin{subfigure}[ht]{0.32\textwidth}
			\includegraphics[width=\textwidth]{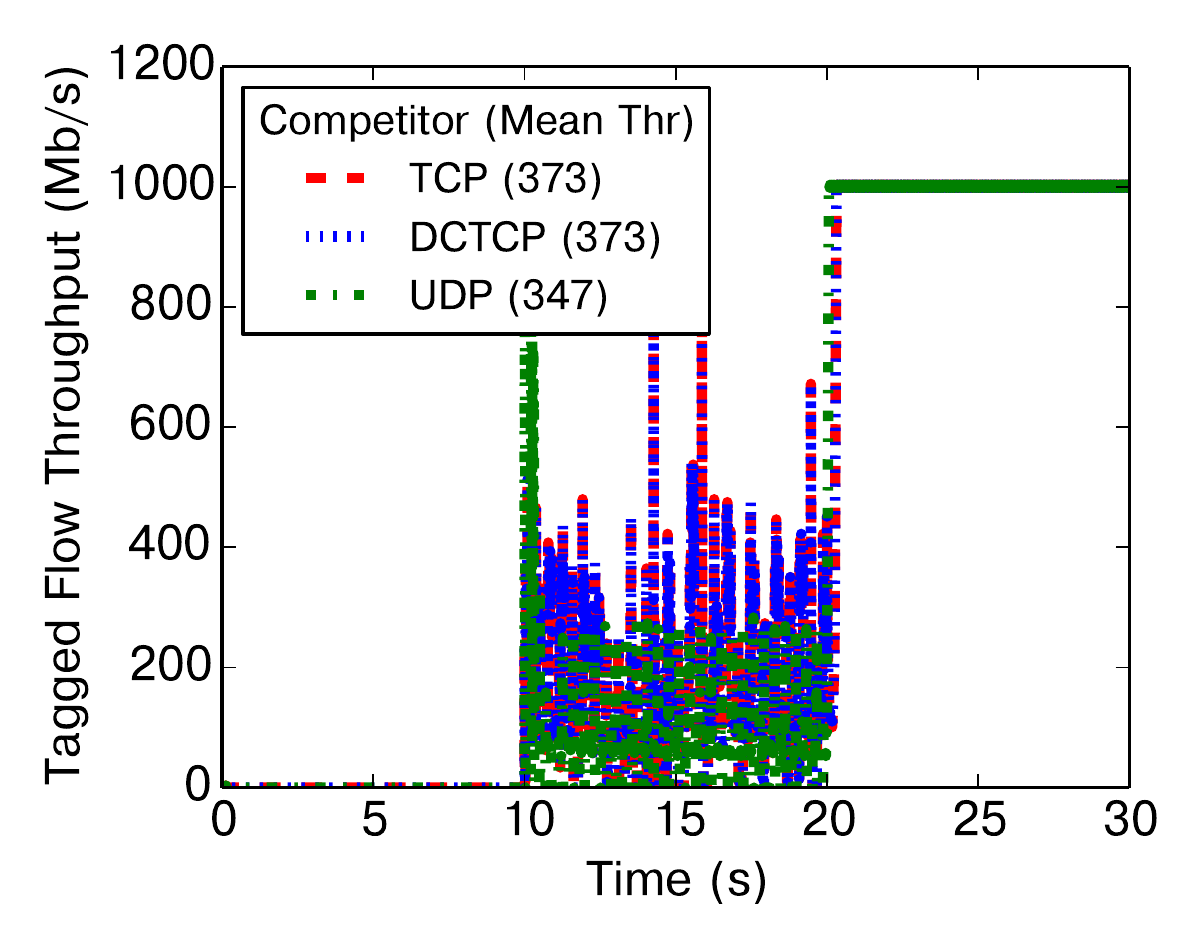}
                \caption{50ms monitor interval}
                \label{fig:50ms}
        \end{subfigure}
				\caption{The instantaneous and mean goodput of the tagged TCP flow while competing with either TCP, DCTCP or UDP using a control period of 1ms (10RTT), 10ms (100RTT) or 50ms (500RTT) is used.}
		\label{fig:multcontrol}
\end{figure*}

In SDN environments, controller delays are also of major concern. To study the sensitivity of the system to the control loop delays, We conduct the previous experiment involving 4 senders however we  vary the control loop delay to a smaller control delay of 10 RTTs and larger delays of 100 RTT and 500 RTT. \figurename~\ref{fig:1ms} shows that to achieve faster convergence smaller switch-controller-hypervisor delay is always preferable. \figurename~\ref{fig:10ms}~and~\ref{fig:50ms} shows that flows oscillations and convergence period increases as the controller delay increases to 10ms and 50ms. This shows that the performance of rate control schemes such as SDN-GCC depends on ensuring that the controller can react within moderate delays.

\section{Testbed implementation of SDN-GCC}
\label{sec:testbed}

\begin{figure}[t]
\captionsetup[subfigure]{justification=centering}
\centering
        \begin{subfigure}[ht]{0.54\columnwidth}
      \includegraphics[width=\textwidth]{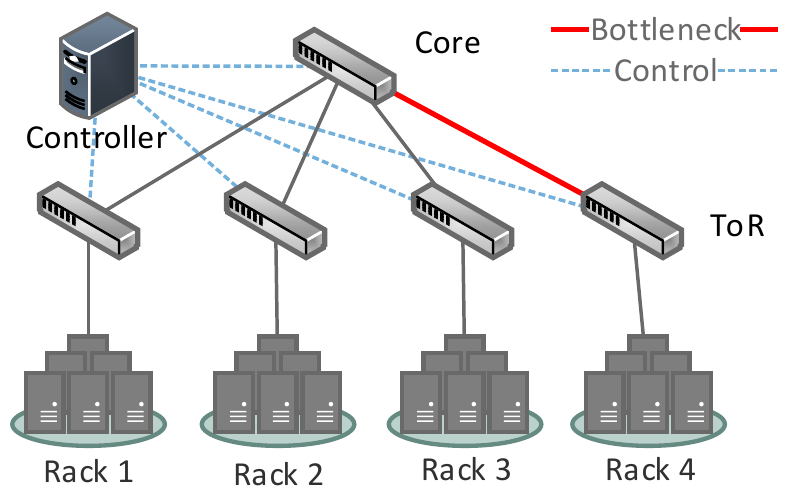}
        \caption{The testbed topology}
        \label{fig:topo}							
        \end{subfigure}
	\quad		
      \begin{subfigure}[ht]{0.38\columnwidth}	
        \includegraphics[width=\textwidth]{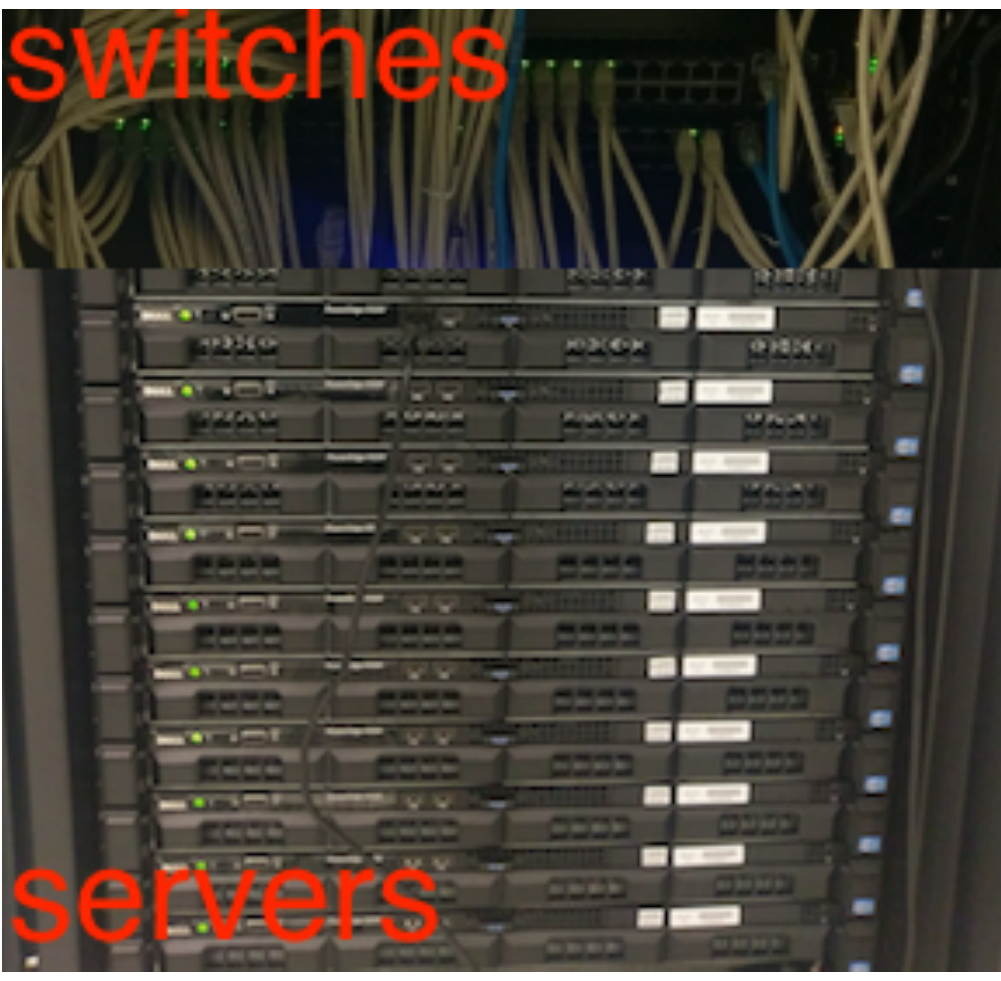}
	 \caption{The actual testbed}
	 \label{fig:acttest}
	 \end{subfigure}
	 \caption{A real testbed for experimenting with SDN-GCC framework}
	\label{fig:SDGenCCtestbed}
\end{figure}

We have implemented SDN-GCC Control application as a separate application program in python for any python-based controller (e.g., Ryu \cite{Ryu} SDN framework in our testbed). Since, most popular cloud management software (e.g., OpenStack) use OpenvSwitch~\cite{OpenvSwitch} as their end-host (hypervisor) networking layer. We implemented SDN-GCC shim-layer as a patch to the Kernel data-path module of OvS. We added the token-bucket rate limiters and the congestion message handler (i.e., the shim-layer) in the packet processing pipeline in the data-path of OvS. In a virtualized environment, OvS forwards the traffic for inter-VM, intra-Host and inter-Host communications. This leads to an easy and straightforward way of deploying the shim-layer at the end-hosts by only applying a patch and recompiling the OvS kernel module, introducing minimal impact on the operations of production DC networks with no need for a complete shutdown. Specifically, deployment can be carried out by the management software responsible for admission and monitoring of the data center.

We set up a testbed as shown in \figurename~\ref{fig:SDGenCCtestbed}. All machines' internal and the outgoing physical ports are connected to the patched OvS. We have 4 racks consisting of 7 servers each (rack 1, 2 and 3 are senders and rack 4 is receiver) all servers are installed with Ubuntu Server 14.04 LTS running kernel v3.16 and are connected to the ToR switch through 1 Gbps links. Similarly, the machines are installed with the iperf \cite{iperf} program for creating elephant flows and the Apache web server hosting a single webpage \textbf{"index.html"} of size 11.5KB for creating mice flows. We setup different scenarios to reproduce both incast and buffer-bloating situations with bottleneck link in the network as shown in \figurename~\ref{fig:SDGenCCtestbed}. Various iperf and/or Apache client/server processes are created and associated with their own virtual ports on the OvS at the end-hosts. This allows for scenarios with large number of flows in the network to emulate a data center with various co-existing applications. The base RTT is $\approx$200 $\mu$s without queuing and $\approx$1 ms with queuing, hence we set the monitoring/sampling interval to conservative values larger than $\approx$1 ms ($\approx$5 times the base RTT). %

\begin{figure*}[t]
\captionsetup[subfigure]{justification=centering}
\centering
			\begin{subfigure}[ht]{0.32\textwidth}
            \includegraphics[width=\textwidth]{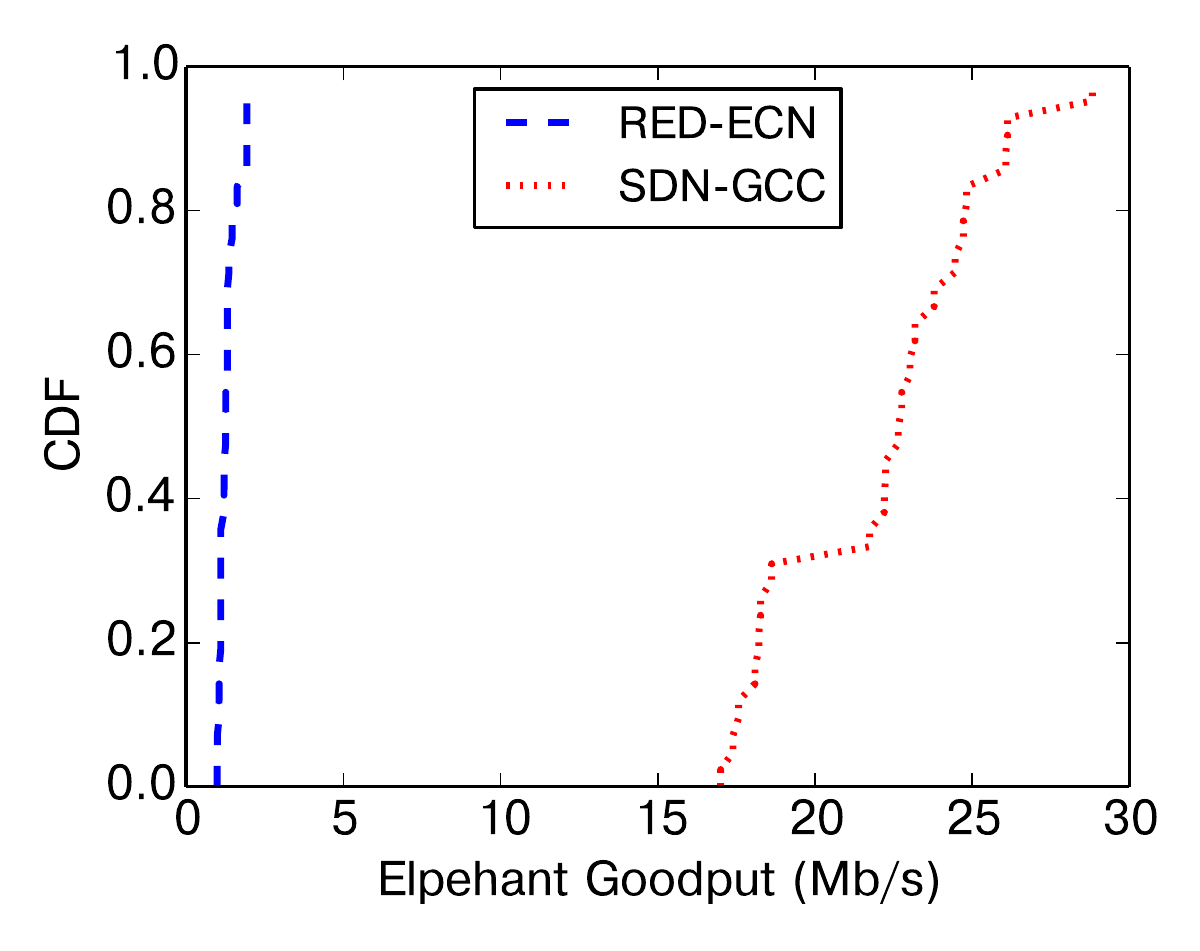}             
						\caption{AVG TCP goodput}
                \label{fig:sdgencc-goodtcp}
       \end{subfigure}
       \hfill
			\begin{subfigure}[ht]{0.32\textwidth}
            \includegraphics[width=\textwidth]{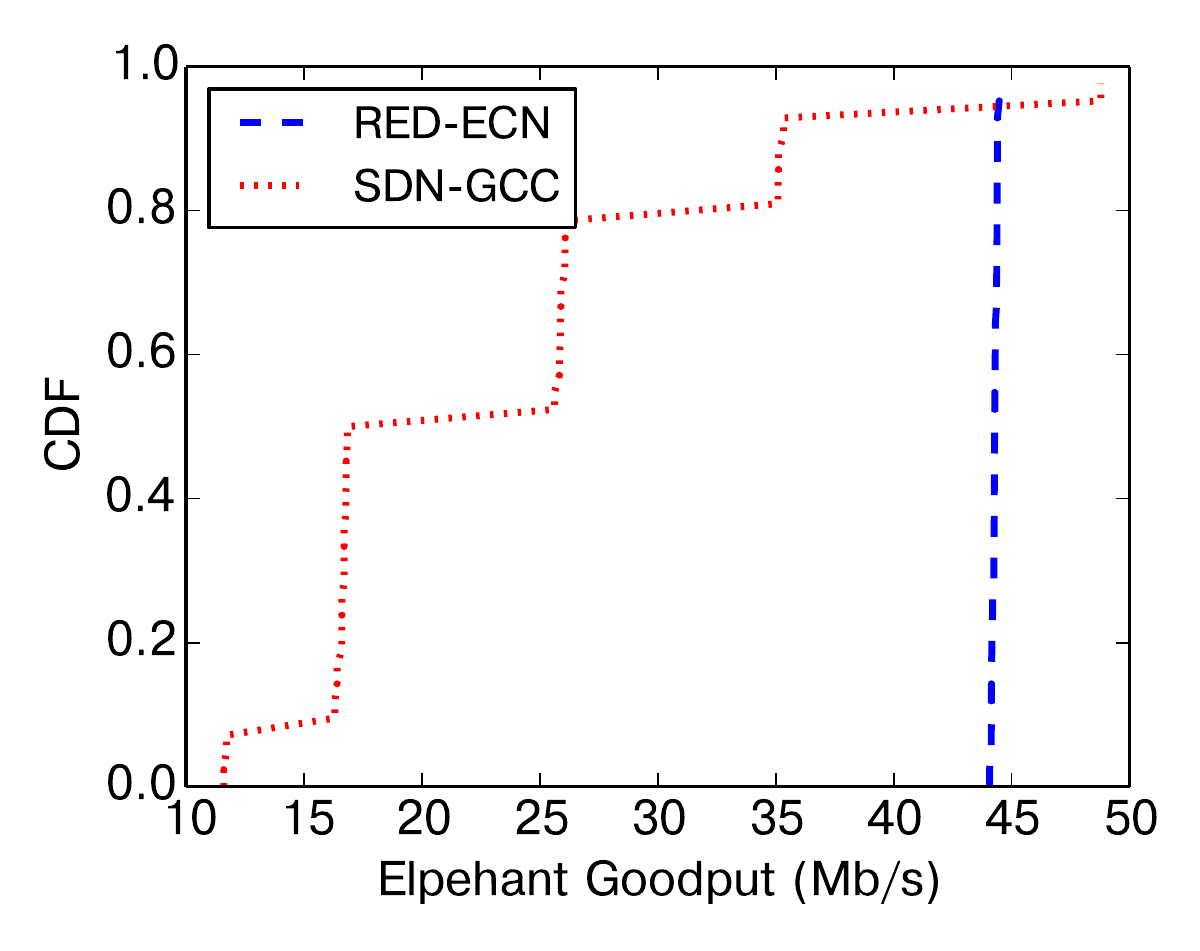}             
						\caption{AVG elephant UDP goodput}
              \label{fig:sdgencc-goodudp}
       \end{subfigure}
				\hfill
			\begin{subfigure}[ht]{0.32\textwidth}
       \includegraphics[width=\textwidth]{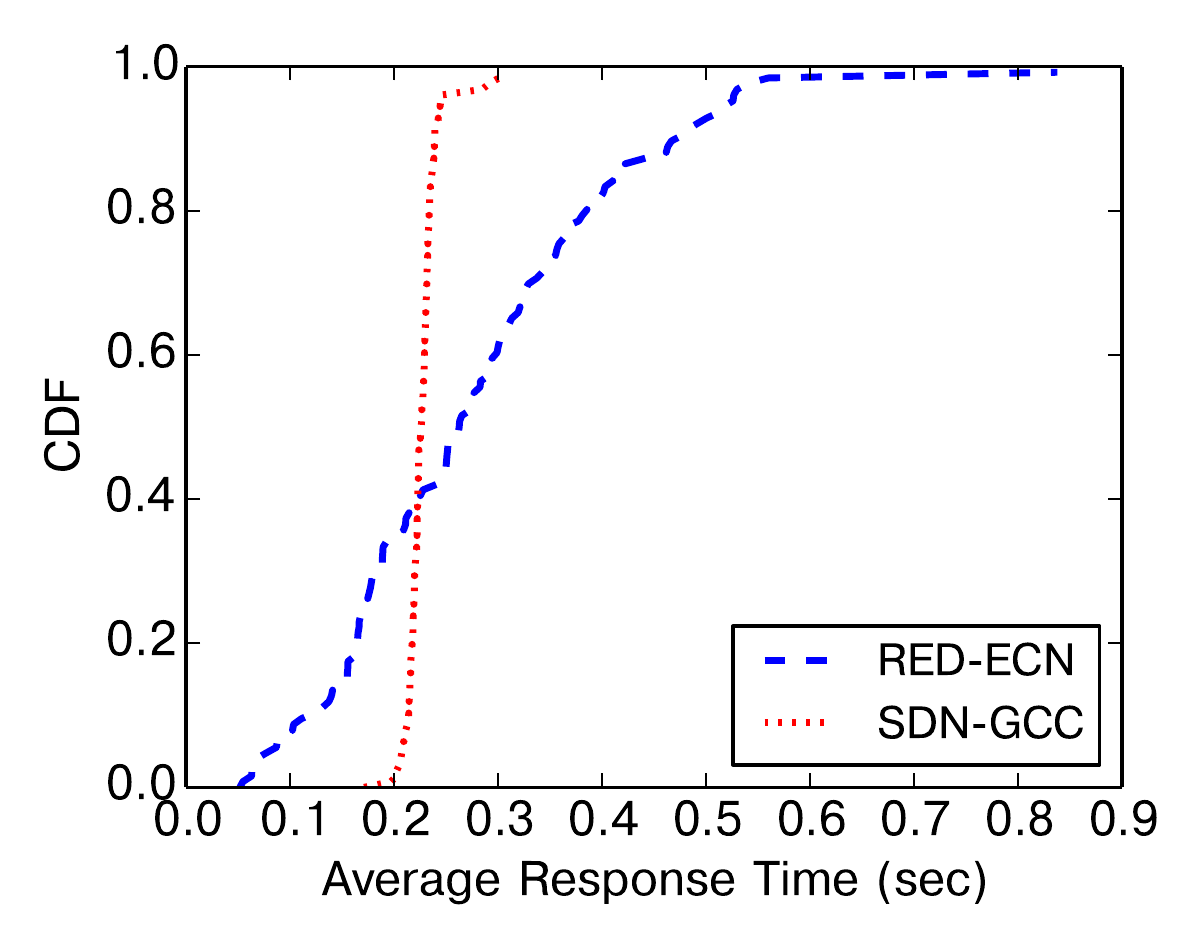}
                 \caption{AVG FCT for mice}
                \label{fig:sdgencc-avg}
                  \end{subfigure}		
				\caption{Testbed experiment involving 126 TCP mice competing against 21 TCP (same variant)  and 21 UDP elephants}
				\label{fig:sdgencctest}
\end{figure*}

We run a scenario in which TCP and UDP elephant flows are competing for bandwidth and to test the agility of SDN-GCC, a burst of mice TCP flows is introduced to compete for bandwidth in a short-period of time. We first generate 7  TCP iperf flows and another 7 UDP iperf flows from each sending rack for 20 secs resulting in 42 ($2 \times 7 \times 3 = 42$) elephants at the bottleneck. At the $10^{th}$sec, we use Apache Benchmark \cite{apacheb} to request \textbf{"index.html"} webpage (10 times) from each of the 7 web servers on each sending rack ($7 \times 6 \times 3 = 126$ in total). \figurename~\ref{fig:sdgencc-goodtcp}~and~\ref{fig:sdgencc-goodudp} show that the TCP elephants are able to grab their share of bandwidth regardless of the existence of non-well-behaved UDP traffic. In addition, \figurename~\ref{fig:sdgencc-avg} suggests that mice flows still benefit from SDN-GCC by achieving a smaller and nearly smooth (equal) flow completion time on average with a smaller standard deviation demonstrating SDN-GCC's effectiveness in apportioning the link capacity. In summary, SDN-GCC effectively tackles congestion and allocates the capacity among various flow types as expected.

To summarize this simulation and experimental study, SDN-GCC seems to be able to efficiently allocate the bandwidth among various flow types and alleviate possible congestion in the network core. 

\section{Related Work}
\label{sec:related}
HyGenICC~\cite{Ahmed-ICC-2016-1} can be comparable or complementary to a number of works on cloud network resource allocation that have been proposed recently. Seawall \cite{Shieh2011} is a system proposed for sharing network bandwidth, it provides per-VM max-min weighted fair share using explicit feedback end-to-end congestion notification based on losses for rate adaptation. Seawall requires modifications to network stack which incurs a large overhead and may interfere with middleboxes operations. SecondNet \cite{Guo2010} is designed to divide network among tenants and enforce rate limits, but is limited to providing static bandwidth reservation between pairs of VMs. Oktopus \cite{Ballani2011a} argues for predictability by enforcing a static hose model using rate limiters. It computes rates using a pseudo-centralized mechanism, where VMs communicate their pairwise bandwidth consumption to a tenant-specific centralized coordinator. This control plane overhead limits reaction times to more than 2 seconds which is inadequate for the fast changing and dynamic traffic nature in datacenters. FairCloud \cite{Popa2012} designs better policies for sharing bandwidth and explored fundamental trade-offs between network utilization, minimum guarantees and payment proportionality, for a number of sharing policies. EyeQ \cite{Jeyakumar2013} provides per-VM max-min weighted fair shares in the context of a full bisection bandwidth datacenter topology where congestion is limited to the first and the last hops. %
By simplifying rate limiters and coupling congestion control to make them dynamic entities rather than static, HyGenICC can achieve similar objectives as these proposals in an easy to deploy manner with minimal CPU and network overhead. HyGenICC is designed to operate with commodity infrastructure and traditional protocols used by current production datacenter/cloud, to be a readily deployable solution. Finally, HyGenICC can leverage the popularity of Open vSwitch (OvS) usage by cloud management frameworks like openstack to implement its mechanism with minor modifications to OvS that do not require any new protocols, software and hardware.

A number of recent proposals implemented different system designs for cloud network resources allocation. ``Seawall" \cite{Shieh2011} is a system designed solely for sharing network capacities by achieving per-VM max-min weighted allocations using explicit end-to-end feedback messaging for rate adaptation. Seawall adds new encapsulation protocol to network stack on top of transport headers which incurs a large processing and messaging overhead as well as rendering it into a non middle-box friendly solution. ``Secondnet" \cite{Guo2010} is a system proposed to divide network among tenants via rate limits enforcements, however, it only supports static bandwidth reservation among tenants' VMs. ``EyeQ" ~\cite{Jeyakumar2013} adopts per-VM max-min weighted fair sharing in the context of a full bisection bandwidth datacenter topology. Its downside is the design assumption that congestion is limited to first and last hops. ``RWNDQ"~\cite{Ahmed-CLOUDNET-2015, Ahmed-IPCCC-2015} is a fair-share allocation AQM for TCP in data centers, however it resolves contention among flows using TCP as their transport. RWNDQ also shares the drawback of low deployment potential which is common to all switch solutions like \cite{Cheng2014, Ahmed-ICC-2016-1, Ahmed-ICC-2016-2}. ``HyGenICC" \cite{Ahmed-ICC-2016-1} is a hypervisor-based IP-based congestion control mechanism that relies on a collaborative information exchange between hypervisors. The solution involves the use of ECN marking as congestion indication which is aggregated and fed back between hypervisors to enable network bandwidth partitioning through dynamic (adaptive) rate limiters. In spite of the appealing performance gains achieved. In general, these mechanisms have some or all of the following drawbacks: 
\begin{enumerate}
\item \textbf{Security}: Introduction of new protocols or using reserved headers may interfere with the operation of the middle-boxes (e.g., HyGenICC uses IP reserved bits known as ``Evil-bit" \cite{EVILBIT} which was used for security testing).
\item \textbf{Overhead}: Flow tracking and feedback packets crafting on a per VM-to-VM basis adds burden to the hypervisor's processing overhead.
\item \textbf{Locality}: The lack of global information about the dynamic network conditions results into hypervisor-based solutions to react only to the perceived VM-to-VM congestion (e.g., transient and non-persistent) rather than the network-wide congestion state.
\item \textbf{Mutli-Path}: VM-to-VM packets are not guaranteed to take the same path when multi-path routing (e.g., ECMP) is used, leading to under estimation of the actual VM-to-VM congestion by the end-points. 
\end{enumerate}
Recently, SDN has seen a growing number of deployments for intra- and inter-data center networks \cite{SDNDC, Feamster2013, Akyildiz2014}. SDN was also invoked to address complex congestion events such as TCP-Incast in data centers~\cite{Ahmed-GLOBECOM-2015,Ahmed-GLOBECOM-2018,Ahmed-LCN-2016,Ahmed-ICC-2017,Ahmed-ANNALS-2017,Ahmed-LCN-2017,Ahmed-ITCE-2019}. Hence, we address the aforementioned drawback of the former schemes by taking advantage of the rich information, flexibility and global scope provided by the SDN framework.  We show that SDN-GCC~\cite{Ahmed-LCN-2017} is a middle-box and mutli-path friendly solution that achieves similar design goals with lesser deployment overhead and lower CPU and network overhead. This is achieved by leveraging simple rate limiters and incorporating a network-aware SDN controller towards building a dynamic adaptive system. The essence of SDN-GCC is to address the increasing trend and shift to SDN infrastructures while keeping traditional transport protocols unchanged in current production data centers.

\section{Conclusion and future work}
\label{sec:conclusion}
In this paper, we set to build a system that relies of the pervasive availability of SDN capable switches in data centers to provide a centralized congestion control mechanism with a small deployment overhead onto production data centers. Our system achieves better bandwidth isolation and improved application performance. SDN-GCC is a SDN framework that can enforce efficient network bandwidth allocation among competing VMs by employing simple building blocks such as rate limiters at the hypervisors along with an efficient SDN application. SDN-GCC is designed to operate with low overhead, on commodity hardware, and with no assumption of tenant's cooperation which makes a great composition for the deployment in SDN-based data center networks. SDN-GCC was shown, via simulation and deployment, to efficiently divide network bandwidth among active VMs, by enforcing the target rates, regardless of the transport protocols in use.

\balance
\small
\def\url#1{}
\bibliographystyle{ieeetr}
\bibliography{paper21,online,mypapers}
\balance
\end{document}